\documentclass[12pt,epsf,amssymb]{article} \textheight =23 truecm
\def\lsim{\raise0.3ex\hbox{$<$\kern-0.75em\raise-1.1ex\hbox{$\sim$}}}
\def\gsim{\raise0.3ex\hbox{$>$\kern-0.75em\raise-1.1ex\hbox{$\sim$}}}
\def\noi{\noindent} \def\nn{\nonumber} \def\bea{\begin{eqnarray}}
\def\eea{\end{eqnarray}} \def\beq{\begin{equation}}
\def\eeq{\end{equation}} 
\def\beeq{\begin{eqnarray}} \def\eeeq{\end{eqnarray}} \def\R{ {\rm R
\kern -.31cm I \kern .15cm}} \def\C{ {\rm C \kern -.15cm \vrule
width.5pt \kern .12cm}} \def\Z{ {\rm Z \kern -.27cm \angle \kern
.02cm}} \def\N{ {\rm N \kern -.26cm \vrule width.4pt \kern .10cm}}
\def\1{{\rm 1\mskip-4.5mu l} }

\usepackage{graphics} \usepackage{graphicx} \usepackage{epsfig}

\begin{document} \begin{center} {\large \bf Subleading form factors
at order 1/m$_{\bf Q}$ in terms of leading quantities using the
non-forward amplitude in HQET} \\

\vskip 1 truecm {\bf F. Jugeau, A. Le Yaouanc, L. Oliver and J.-C.
Raynal}\\

{\it Laboratoire de Physique Th\'eorique}\footnote{Unit\'e Mixte de
Recherche UMR 8627 - CNRS }\\    {\it Universit\'e de Paris XI,
B\^atiment 210, 91405 Orsay Cedex, France} \end{center}

\vskip 1 truecm

\begin{abstract} We consider the non-forward amplitude within the Heavy
Quark Effective Theory. We show that one can obtain new information on
the subleading corrections in $1/m_Q$. We illustrate the method by
deriving new simple relations between the {\it functions} $\xi_3(w)$
and $\overline{\Lambda}\xi (w)$ and the sums $\sum\limits_n \Delta
E_j^{(n)} \tau_j^{(n)}(1) \tau_j^{(n)}(w)$ ($j = {1\over 2}, {3 \over
2})$, that involve leading quantities, namely the Isgur-Wise functions
$\tau_j^{(n)}(w)$ and the level spacings $\Delta E_j^{(n)}$. Our
results follow because the non-forward amplitude $B(v_i) \to
D^{(n)}(v') \to B(v_f)$ depends on three variables $(w_i, w_f, w_{if})
= (v_i\cdot v', v_f \cdot v', v_i \cdot v_f)$ independent in a certain
domain, and we consider the zero recoil frontier $(w,1,w)$ where only a
finite number of $J^P$ states contribute $\left ( {1\over 2}^+, {3
\over 2}^+\right )$. These sum rules reduce to known results at $w =
1$, for $\overline{\Lambda}$ obtained by Voloshin, and for $\xi_3(1)$
obtained by Le Yaouanc et al. and by Uraltsev, and generalizes them to
all values of $w$. We discuss phenomenological applications of these
results, in particular the check of Bakamjian-Thomas quarks models and
the comparison with the QCD Sum Rules approach. \end{abstract}

\vskip 2 truecm

\noi LPT Orsay 04-20 \par \noi May 2004\par \vskip 1 truecm

\noindent e-mails : frederic.jugeau@th.u-psud.fr,
leyaouan@th.u-psud.fr, oliver@th.u-psud.fr \newpage \pagestyle{plain}

\section{Introduction.} \hspace*{\parindent} In the leading order of
the heavy quark expansion of QCD, Bjorken sum rule (SR) \cite{1r,2r}
relates the slope of the elastic Isgur-Wise (IW) function $\xi (w)$, to
the IW functions of the transition between the ground state $j^P = {1
\over 2}^-$ and the $j^P = {1 \over 2}^+, {3 \over 2}^+$ excited
states, $\tau_{1/2}^{(n)}(w)$, $\tau_{3/2}^{(n)}(w)$ at zero recoil $w
= 1$ ($n$ is a radial quantum number). This SR leads to the lower bound
$- \xi '(1) = \rho^2 \geq {1 \over 4}$. A new SR was formulated by
Uraltsev in the heavy quark limit \cite{3r}, involving also
$\tau_{1/2}^{(n)}(w)$, $\tau_{3/2}^{(n)}(w)$, that implies, combined
with Bjorken SR, the much stronger lower bound $\rho^2 \geq {3 \over
4}$. A basic ingredient in deriving this bound is the consideration of
the non-forward amplitude $B(v_i) \to D^{(n)}(v') \to B(v_f)$, allowing
for general $v_i$, $v_f$, $v'$ and where $B$ is a ground state meson.
In refs. \cite{4r,5r,6r} we have developed, in the heavy quark limit of
QCD, a manifestly covariant formalism within the Operator Product
Expansion (OPE), using the matrix representation for the whole tower of
heavy meson states \cite{7new}. We did recover Uraltsev SR plus a
general class of
SR that allow to bound also higher derivatives of the IW function. In
particular, we found two bounds for the curvature $\xi ''(1) =
\sigma^2$ in terms of $\rho^2$, that imply $\sigma^2 \geq {15 \over
16}$. \par

The object of the present paper is to extend the formalism to IW
functions at subleading order in $1/m_Q$. \par

The general SR obtained from the OPE can be written in the compact way
\cite{4r}
\beq \label{1e} L_{Hadrons}(w_i,w_f,w_{if}) = R_{OPE}(w_i,w_f,w_{if})
\eeq

\noi where the l.h.s. is the sum over the intermediate $D$ states,
while the r.h.s. is the OPE counterpart. Using the trace formalism
\cite{7r}, this expression writes, in the heavy quark limit \cite{4r}~:
$$\sum_{D=P,V} \ \sum_n {\rm Tr} \left [
\overline{B}_f(v_f)\Gamma_f D^{(n)}(v')\right ] \ {\rm Tr} \left [
\overline{D}(v') \Gamma_i B_i (v_i)\right ] \xi^{(n)}(w_i)
\xi^{(n)}(w_f) + \hbox{Excited states}$$
\beq \label{2e} = - 2 \xi (w_{if})\ {\rm Tr}\left [
\overline{B}_f(v_f)\Gamma_f P'_+ \Gamma_i B_i(v_i)\right ]\eeq

\noi where \beq \label{3e} w_i = v_i \cdot v' \qquad w_f = v_f \cdot v'
\qquad w_{if} = v_i \cdot v_f \eeq

\noi and
\beq
\label{4w}
P'_+ = {1 + {/\hskip - 2 truemm v}' \over 2}
\eeq

\noi is the positive
energy projector on the intermediate $c$ quark, and we assume that the
IW functions are real. The ground state $B$ meson can be either a
pseudoscalar or a vector, the doublet ${1 \over 2}^-$. The heavy quark
currents considered in the previous expression are
\beq \label{4e} \overline{h}_{v'}\Gamma_i h_{v_i} \qquad \qquad
\overline{h}_{v_f}\Gamma_f h_{v'} \eeq

\noi $B(v)$, $D(v)$ are the $4 \times 4$ matrices representing the $B$,
$D$ states \cite{7r}, and $\overline{B} = \gamma^0B^+\gamma^0$ denotes
the Dirac conjugate matrix. The domain for the variables
$(w_i,w_f,w_{if})$ is \cite{4r}~:
$$w_i \geq 1 \qquad \qquad w_f \geq 1$$
\beq \label{5e} w_iw_f - \sqrt{(w_i^2 - 1) (w_f^2 - 1)} \leq w_{if}
\leq w_iw_f + \sqrt{(w_i^2 - 1)(w_f^2 -1)} \ . \eeq

\noi We will now consider $1/m_Q$ corrections to the heavy quark limit
SR (\ref{2e}).

The paper is organized as follows. In Section 2 we set the problem of
obtaining sum rules involving subleading quantities in $1/m_Q$ within
the OPE. In Section 3 we make explicit the formalism of the corrections
in $1/m_b$, the $b$-quark being the external quark, using the
formalisms of Falk and
Neubert \cite{8r} and of Leibovich et al. \cite{9r} to
parametrize the ${1\over 2}^- \to {1\over 2}^-$ and ${1\over 2}^- \to
{1\over 2}^+, {3\over 2}^+$ form factors, that we extend to the
${1\over 2}^- \to {3\over 2}^-$ transitions. In Sections 4 and 5 we
write down the SR obtained respectively if the initial and final meson
is a pseudoscalar $B$ or a vector $B^*$. We factorize polynomials in the
variables $(w_i, w_f, w_{if})$ that allow to obtain simple results
going to the interesting frontier $(w, 1,w)$ of the domain
(\ref{5e}). We generalize our
results to any ${1\over 2}^- \to j^P$ transition. From the obtained SR
we get enough information to write our fundamental results for the
subleading quantities $\overline{\Lambda}\xi (w)$ and $\xi_3(w)$ in
terms of leading quantities in Section 6. In Section 7 we use as input
the results of the Bakamjian-Thomas class of quark models --
that satisfy all the necessary properties in the heavy quark limit -- to
obtain phenomenologically useful results, that appear to be consistent,
in our quite different approach, with the QCD Sum Rules. Finally, in
Section 8 we conclude and set up the
program that remains to be pursued. In Appendix A we demonstrate the
identity between two subleading parameters defined by Falk and Neubert
\cite{8r} and by Uraltsev \cite{3r}. In Appendix B we discuss the
experimental situation of the leading $P$-wave IW functions
$\tau_{1/2}^{(0)}(w)$
and $\tau_{3/2}^{(0)}(w)$.

\section{The Operator Product Expansion and the\break \noindent
corrections at first order in 1/m$_{\bf Q}$.} \hspace*{\parindent} Our
starting point \cite{12r} is the $T$-product
\beq \label{6a} T_{fi}(q) = i \int d^4x\ e^{-iq\cdot x} \
<B(p_f)|T[J_f(0)J_i(x) ]| B(p_i)> \eeq

\noi where $J_f(x)$, $J_i(y)$ are the currents (the convenient notation
for the subindices $i$, $f$ will appear clear below)~:
\beq \label{7a} J_f(x) = \overline{b}(x) \Gamma_f c(x) \qquad \qquad
J_i(y) = \overline{c}(y)\Gamma_i b(y) \eeq

\noi and $p_i$ is in general different from $p_f$. \par

Inserting in this expression hadronic intermediate states, $x^0 < 0$
receives contributions from the direct channel with hadrons with a
single heavy quark $c$, while $x^0 > 0$ receives contributions from
hadrons with $b\overline{c}b$ quarks, the $Z$ diagrams~: \bea
\label{8a} &&T_{fi}(q) =  \sum_{X_c} (2 \pi)^3
\ \delta^3( {\bf q} + {\bf p}_i - {\bf p}_{X_c})
{<B_f|J_f(0)|X_c> <X_c|J_i(0)|B_i> \over q^0 + E_i - E_{X_c} + i 
\varepsilon}\nn \\ &&-
\sum_{X_{\overline{c}bb}} (2\pi)^3 \ \delta^3 ( {\bf q} - {\bf p}_f
+ {\bf p}_{X_{\overline{c}bb}}) {<B_f|J_i(0)|X_{\overline{c}bb}>
<X_{\overline{c}bb}|J_f(0)|B_i> \over q^0 - E_f + 
E_{X_{\overline{c}bb}}- i \varepsilon}
\ . \eea

We will consider the following limit
\beq
\label{no4}
m_c \gg m_b \gg \Lambda_{QCD}\ .
\eeq

\noi The difference between the two energy denominators is large
\beq
\label{no5}
    q^0 - E_f + E_{X_{\overline{c}bb}} - \left ( q^0 + E_i - E_{X_c}
\right ) \sim 2m_c\ .
\eeq

\noi Therefore, we can in this limit neglect the second term, and we
will consider the imaginary part of the direct diagram, the first term
in (\ref{8a}), the piece proportional to
\beq
\label{no6}
\delta \left ( q^0 + E_i - E_{X_c}\right )\ .
\eeq

\noi One can see this point otherwise. The two cuts corresponding to
the two terms in (\ref{8a}) are widely separated, and one can isolate the
imaginary part of the first term by a suitable integration contour in
the $q^0$ complex plane. Notice that one can choose $q^0$ such that
there is a left-hand cut, even in the conditions (\ref{no4}). This
means that $q^0$ is of the order of $m_c$ and $m_c-q^0$ is fixed, of
the order $m_b$. Our conditions are, in short, as follows~:
\beq
\label{no6bis}
\Lambda_{QCD} \ll m_b \sim m_c - q^0 \ll q^0 \sim m_c \ ,
\eeq

\noi consistent with (\ref{no6}). To summarize, we are considering
the heavy quark limit for the $c$
quark, but we allow for a large finite mass for the $b$ quark. \par

Unlike the case of the forward amplitude $(p_i = p_f)$, the imaginary
part of the direct diagram in (\ref{8a}) will not be related to a
positive definite absorptive part, because we are in the more general
case of the non-forward amplitude. However, we are allowed to consider
this imaginary part.\par

In the conditions (\ref{no6bis}), or choosing the suitable integration
contour \cite{10r} \cite{11r}, we can write therefore, integrating over
$q^0$
\beq
\label{no7}
T^{abs}_{fi}({\bf q}) \cong \sum_{X_c} (2 \pi)^3 \ \delta^3\left
({\bf q} + {\bf p}_i - {\bf q}_{X_c}\right ) \ <B_f|J_f(0)|X_c>\
<X_c|J_i(0)|B_i>\ .
\eeq

\noi Finally, integrating over ${\bf q}_{X_c}$ and defining $v' = {q
+ p_i \over m_c}$ one gets
\beq
\label{no8}
T^{abs}_{fi} \cong \sum_{D_n} <B_f(v_f)|J_f(0)|D_n(v')>\
<D_n(v')|J_i(0)|B_i(v_i)>
\eeq

\noi where we have denoted by $D_n(v')$ the charmed intermediate states.\par

The $T$-product matrix element $T_{fi}(q)$ (\ref{6a}) is given,
alternatively, in terms of quarks and gluons, by the expression
\beq
\label{no9}
T_{fi} (q) = - \int d^4 x \ e^{-iq\cdot x} \ <
B(p_f)|\overline{b}(0)\Gamma_f S_c (0, x) \Gamma_i b(x) |B(p_i)>
\eeq

\noi where $S_c (0, x)$ is the $c$ quark propagator in the background
of the soft gluon field \cite{14new}. \par

Since we are considering the absorptive part in the $c$ heavy quark
limit of the direct graph in (\ref{8a}), this quantity can be then
identified with (\ref{no9}) where $S_c(x,0)$ is replaced by the
following expression \cite{GK}
\beq
\label{no10}
S_c(0, x) \to e^{im_cv'\cdot x} \ \Phi_{v'}[0, x] D_{v'}(x)
\eeq

\noi where $D_{v'}(x)$ is the {\it cut} free propagator of a heavy quark
\beq
\label{no11}
D_{v'}(x) = P'_+ \int {d^4k \over (2 \pi)^4} \delta (k\cdot v')
e^{ik\cdot x} = P'_+ \int_{-\infty}^{\infty} {dt \over 2\pi} \delta^4
(x- v't)
\eeq

\noi with the positive energy projector defined by
\beq
\label{no12}
P'_+ = {1 \over 2} (1 + {/ \hskip - 2 truemm v}')\ .
\eeq

The eikonal phase $\Phi_{v'}[0, x]$ in (\ref{no10}) corresponds to
the propagation of the $c$ quark from the point $x = v't$ to the
point 0, that is given by
\beq
\label{no13}
\Phi_{v'} [0, v't] = P \exp \left ( - i \int_0^t ds\ v'\cdot A(v's)\right )\ .
\eeq

\noi This quantity takes care of the dynamics of the soft gluons in
HQET along the classical path $x = v't$. \par

Inserting (\ref{no10})-(\ref{no12}) into (\ref{no9}) we obtain
\bea
\label{no15}
&&T^{abs}_{fi}(q) = \int d^4 x \ e^{-i(q-m_c v')\cdot x}
\int_{-\infty}^{\infty} {dt \over 2 \pi} \delta^{4} (x - v't)\nn \\
&&<B(p_f) | \overline{b}(0) \Gamma_f
P'_+ \Phi_{v'} [0, x] \Gamma_i b(x) |B(p_i)>\ + \ O(1/m_c)  \ .
\eea

\noi Integrating over $x$ in (\ref{no15}) and making explicit (\ref{no13}),
\bea
\label{no16}
&&T^{abs}_{fi}(q) = \int_{-\infty}^{\infty} {dt \over 2\pi} \
e^{-i(q-m_c v')\cdot v't} \nn \\
&&<B(p_f) | \overline{b}(0) \Gamma_f
P'_+ P \exp\left ( - i \int_0^t ds\ v'\cdot A(v's)\right ) \Gamma_i
b(v't) |B(p_i)>\nn\\
&& + \ O(1/m_c)  \ .
\eea

Performing first the integration over $q^0$ one obtains simply $\delta
(v'^0t)$, that forces $t= 0$, and making the trivial integration over
$t$ one obtains finally the OPE matrix element that must be identified
with (\ref{no8})~:
\bea
\label{no17}
&&T^{abs}_{fi} \cong \ <B(p_f) | \overline{b}(0) \Gamma_f {1 +
{/\hskip -2 truemm v}' \over 2v'^0} \Gamma_i b(0)|B(p_i)>\ + \
O(1/m_c)  \ .
\eea

Therefore, we end up with the sum rule
\bea
\label{24NR}
&&\sum_{D_n} <B_f(v_f)|J_f(0)|D_n(v')>\ <D_n(v')|J_i(0)|B_i(v_i)>\nn \\
&&=\ <B(v_f) |\overline{b}(0)\Gamma_f {1 + {/\hskip -2 truemm v}'
\over 2v'^0} \Gamma_i b(0)|B(v_i)>\ + \ O(1/m_c)
\eea

\noi that is valid for {\it all powers} of an expansion in $1/m_b$,
but only to leading order in $1/m_c$. \par

On the other hand, making use of the HQET equations of motion, the
field $b(x)$ in (\ref{24NR}) can be decomposed into upper and lower
components as follows \cite{N}
\beq
\label{no14}
b(x) = e^{-im_bv\cdot x} \left ( 1 + {1 \over 2m_b + iv\cdot
\overrightarrow{D}} \ i\overrightarrow{/\hskip - 3 truemm D}\right )
h_{v}(x)
\eeq

\noi where the second term corresponds to the lower components and
can be expanded in a series in powers of $D_{\mu}/m_b$, and $v$ is an
arbitrary four-velocity.\par

Taking into account the normalization of the states in the trace
formalism and the sign convention for the matrix elements, we recover,
in the heavy quark limit (neglecting the term in $1/2m_b)$, the master
formula (2) obtained in ref. \cite{4r}. \par

Including the first order in $1/m_b$, the sum rule reads
\bea
\label{no18}
&&\sum_{D_n} <B_f(v_f)|J_f(0)|D_n(v')>\ <D_n (v')|J_i(0)|B_i(v_i)>\nn \\
&&= \ <B(p_f) |\overline{h}_{v_f}(0)\Gamma_f{1 + {/\hskip -2 truemm
v}' \over 2v'^0} \Gamma_i h_{v_i}(0)|B(p_i)>\nn \\
&&+{1 \over 2 m_b} \ <B(p_f)| \overline{h}_{v_f}(0)\Big [
(-i\overleftarrow{/\hskip - 3 truemm D}) \Gamma_f{1 + {/\hskip -2
truemm v}' \over 2v'^0}\Gamma_i
+ \Gamma_f{1 + {/\hskip -2 truemm v}' \over 2v'^0}\Gamma_i
(i\overrightarrow{/\hskip - 3 truemm D})\Big ] h_{v_i}(0)|B(p_i)>\nn\\
&&+ \ O(1/m_c) + O(1/m_b^2) \ .
\eea

Therefore, in the OPE side we have, besides the leading dimension 3
operator

\beq \label{18a} O^{(3)} = \overline{h}_{v_f} \Gamma_f P'_+ \Gamma_i
h_{v_i} \eeq

\noi the dimension 4 operator
\beq \label{19a} O^{(4)} = \overline{h}_{v_f} \left [ (- i
\overleftarrow{/\hskip - 3 truemm D}) \Gamma_f P'_+ \Gamma_i + \Gamma_f
P'_+ \Gamma_i (i \overrightarrow{/\hskip - 3 truemm D})\right ]
h_{v_i} \ . \eeq

In the SR we have to compute the l.h.s. including terms of order
$1/2m_b$. These terms have been parametrized by Falk and Neubert for
the ${1 \over 2}^-$ doublet and by Leibovich et al. for the transitions
between the ground state ${1 \over 2}^-$ and the ${1 \over 2}^+$, ${3
\over 2}^+$ excited states. These $1/m_b$ corrections are of two
classes~: perturbations of the current, and perturbations of the
Lagrangian (kinetic and magnetic). This will be our guideline to
compute the l.h.s. of the SR, although we will consider the whole tower
of excited states. \par

A remark is in order here, that was already made in ref. \cite{12r}.
Had we taken higher moments of the form $\int dq^0(q^0)^n
T_{fi}^{abs}(q^0)$ ($n > 0$), instead of the lowest one $n = 0$, the
integration over $q^0$ that leads to the simple sum rules
(\ref{24NR}) or (\ref{no18}) would involve higher dimension operators, giving
a whole tower of sum rules \cite{17NR}, \cite{GK}, even in the {\it
leading} heavy quark limit. Our point of view in this paper is
different. We consider the lowest moment $n = 0$, while we expand in
powers of $1/m_b$, keeping the first order in this parameter.\par

Concerning the OPE side in (\ref{no18}), the dimension 4 operator
$O^{(4)}$ (\ref{19a}) is nothing
else but the $1/m_b$ perturbation of the heavy current $O^{(3)} =
\overline{h}_{v_f}\Gamma_fP'_+\Gamma_i h_{v_i}$ since this operator,
containing the Dirac matrix $\Gamma_f P'_+ \Gamma_i$ between heavy
quark fields, can be considered
as a heavy quark current. Indeed, following Falk and Neubert, the
$1/m_b$ perturbation of any
heavy quark current $\overline{h}_{vf}\Gamma h_{vi}$ is given by
\beq \label{20a} \overline{h}_{v_f}\left ( -{i \overleftarrow{/\hskip -
3 truemm D} \over 2m_b}\right ) \Gamma h_{v_i} + \overline{h}_{v_f}\Gamma\left
( {i \overrightarrow{/\hskip - 3 truemm D} \over 2m_b}\right )
h_{v_i} \ . \eeq

However, this perturbation of the current does not exhaust all
perturbations in $1/m_b$. Indeed, we need also to compute the
perturbation of the initial and final wave functions $|B_i(v_i)>$,
$|B_f(v_f)>$ due to the kinetic and magnetic perturbations of the
Lagrangian. This can be done easily following also the prescriptions of
Falk and Neubert to compute these corrections in $1/m_b$ for the leading
matrix element $<B_f(v_f)|\overline{h}_{vf}\Gamma_f P'_+ \Gamma_i
h_{vi}|B_i(v_i)>$, as we will see below.

\section{Setting the formalism for the calculation of the corrections
in 1/m$_{\bf b}$.} \hspace*{\parindent} Considering $B$ or $B^*$
initial and final mesons, we can perturb the SR (\ref{2e}) by $1/m_c$
and $1/m_b$ terms. The perturbation of the r.h.s. is parametrized by
six new subleading IW functions concerning the ground state ${1 \over
2}^-$, denoted by $L_i(w)$ ($i = 1, \cdots 6$), in the notation of Falk
and Neubert \cite{8r}.\par

As for the l.h.s., considering for the moment as intermediate $D$
states the multiplets ${1 \over 2}^-$, ${1 \over 2}^+$, ${3 \over
2}^+$, we have three types of matrix elements
\bea \label{6e} &&<D\left ( {\scriptstyle {1 \over 2}^-}\right
)(v')|\overline{c}\Gamma b|B(v)> \nn \\ &&<D \left ( {\scriptstyle{1
\over 2}^+}\right )(v')|\overline{c}\Gamma b|B(v)> \nn \\ &&<D \left
({\scriptstyle {3 \over 2}^+}\right )(v')|\overline{c}\Gamma b|B(v)> \
. \eea

\noi The corrections in $1/m_b$ or $1/m_c$ to the first matrix element
are given by the same ground state subleading IW functions $L_i(w)$ ($i
= 1, \cdots 6$), while the $O(1/m_b)$ and $O(1/m_c)$ corrections to the
matrix elements $B \to D\left ( {1 \over 2}^+\right )$, $D\left ( {3
\over 2}^+\right )$ have been carefully studied by Leibovich, Ligeti,
Steward and Wise \cite{9r}, and result in a number of new subleading IW
functions. All these corrections are of two types, perturbations of the
heavy quark current, and perturbations of the Lagrangian. \par

Moreover, since, as pointed out by Leibovich et al. [10, Section VI],
the states $D\left ( {3 \over 2}^-, 1^-\right )$ contribute also to
zero recoil at order $1/m_Q$, we will consider the contribution of the
matrix elements
\beq \label{7e} <D\left ( {\scriptstyle {3 \over 2}^-, J^-}\right
)(v')|\overline{c}\Gamma b|B(v)> \qquad (J = 1,2)\ . \eeq

We will show that these contributions do not spoil the simple result
presented below, that can be expressed only in terms of the leading IW
functions $\tau_{1/2}(w)$ and $\tau_{3/2}(w)$. We argue also that
higher $j^P$ intermediate states do not contribute.\par

Let us again underline that we will not take into account radiative hard
gluon corrections, as computed in \cite{10r} for Bjorken SR, in
\cite{3r} for Uraltsev SR and in \cite{11r} for our SR concerning the
curvature of the IW function \cite{6r}. \par

We begin with the general SR in the heavy quark limit (\ref{2e}) and
perturb the heavy quark limit matrix elements with $1/m_c$ and $1/m_b$
corrections. The general expression could then be written, making
explicit the leading and the $1/m_c$ and $1/m_b$ parts~:
\bea \label{8e} &&G_0(w_i, w_f, w_{if}) + E_0(w_i, w_f, w_{if}) + {1
\over 2m_b} \left [ G_b(w_i, w_f, w_{if}) + E_b(w_i, w_f, w_{if})\right
]\nn \\ &&+ {1 \over 2m_c} \left [ G_c(w_i, w_f, w_{if}) + E_c(w_i,
w_f, w_{if})\right ] \nn \\ &&= R_0(w_i, w_f, w_{if}) + {1 \over 2m_b}
R_b(w_i, w_f, w_{if}) + {1 \over 2m_c} R_c(w_i, w_f, w_{if}) \eea

\noi where the subindex $0$ means the heavy quark limit, while the
subindex $b$ or $c$ correspond to the subleading corrections in $1/m_b$
or $1/m_c$, and $G$ or $E$ mean, respectively, ground state or excited
state contributions.\par

In the heavy quark limit, one has
\beq \label{9e} G_0(w_i, w_f, w_{if}) + E_0(w_i, w_f, w_{if}) =
R_0(w_i, w_f, w_{if}) \eeq

\noi that leads to equation (\ref{2e}) and to the results quoted
above \cite{1r}-\cite{6r}.\par

In expression (\ref{8e}) we can vary $m_b$ and $m_c$ as independent
parameters and obtain new SR for the subleading quantities.\par

To obtain information on the $1/m_b$ corrections, it is relatively
simple to proceed as follows. We will assume the formal limit of Section 2~:
\beq \label{10e} m_c \gg m_b \gg \Lambda_{QCD} \eeq

\noi and perturb both sides of the SR (\ref{9e}) by $1/m_b$ terms.
This heuristic procedure gives the same results as the method
demonstrated in Section 2.\par

In this limit, since the parameter $1/m_b$ can be varied at will, one
obtains the relation
\beq \label{11e} G_b(w_i, w_f, w_{if}) + E_b(w_i, w_f, w_{if}) =
R_b(w_i, w_f, w_{if})\ . \eeq

One can compute $G_b(w_i, w_f, w_{if})$ and $E_b(w_i, w_f, w_{if})$
using respectively the formalism of Falk and Neubert \cite{8r} and the
one of Leibovich et al. \cite{9r}, and obtain SR for the different
subleading IW functions $L_i(w)$ $(i = 1, \cdots 6)$. \par

Of course, one can obtain SR by taking the opposite limit $m_b \gg
m_c$, that must be consistent with the preceding ones. In ref.
\cite{12r} we did adopt the Shifman-Voloshin limit \cite{13new}
$m_b, m_c \gg m_b - m_c \gg \Lambda_{QCD}$ for the forward amplitude.\\

To be explicit, let us define these functions from the current matrix
elements, following the notation of Falk and Neubert \cite{8r}~:
\bea \label{12e} &&<D(v')|\overline{Q}'\Gamma Q|B(v)>\ \cong - \xi (w)
\ {\rm Tr} \left [ \overline{D}(v') \Gamma B(v) \right ]\nn \\ &&- {1
\over 2m_b} \ {\rm Tr} \left \{ \overline{D}(v') \Gamma\left [
P_{+}L_{+}(v,v') + P_{-} L_{-}(v,v')\right ] \right \} \eea

\noi in the formal limit $m_c \gg m_b$ (\ref{10e}) that we adopt here.
\par

The $4 \times 4$ matrices write, respectively, for pseudoscalar and 
vector mesons~:
\bea
&&M(v) = P_+(v) (-\gamma_5) \nn \\
&&M(v) = P_+ (v) \ {/\hskip - 2 truemm \varepsilon}_v
\eea

\noi while the subleading $1/m_b$ functions are for pseudoscalar and 
vector mesons~:
\beq \label{13e} P_+(v)
L_+(v,v') + P_-(v) L_-(v,v') = \left [ L_1(w)P_+(v) + L_4(w) P_-(v)
\right ] (- \gamma_5) \eeq
\bea \label{14e} &&P_+(v) L_+(v,v') + P_-(v) L_-(v,v') =\nn \\
&&P_+(v) \left [ {/\hskip - 2 truemm \varepsilon}_{v} L_2(w) + \left
( \varepsilon_v \cdot v'\right ) L_3(w) \right ] + P_-(v) \left [
{/\hskip - 2 truemm \varepsilon}_{v} L_5(w) + \left (
\varepsilon_{v}\cdot v'\right ) L_6(w)\right ]  \eea

\noi where $w = v \cdot v'$. \par

The matrix elements to excited states write \cite{9r}
\bea \label{15e} &&<D\left ( {\scriptstyle {3 \over 2}^+}\right
)(v')|\overline{c}\Gamma b|B(v)>\ \cong  \sqrt{3}\ \tau_{3/2}(w) \ {\rm
Tr} \left [ v_{\sigma} \overline{D}^{\sigma}(v')\Gamma B(v)\right ] \nn
\\ &&+ \ {1 \over 2 m_b} \Big \{ {\rm Tr}\left [ S_{\sigma
\lambda}^{(b)}\overline{D}^{\sigma}(v')\Gamma
\gamma^{\lambda}B(v)\right ] + \eta_{ke}^{(b)}\ {\rm Tr} \left [
v_{\sigma} \overline{D}^{\sigma}(v')\Gamma B(v)\right ]\nn \\ &&+ \
{\rm Tr}\left [ R_{\sigma\alpha\beta}^{(b)}
\overline{D}^{\sigma}(v')\Gamma P_+(v) i \sigma^{\alpha\beta}B(v)\right
]\Big \}\nn \\ && \nn \\ &&<D\left ( {\scriptstyle {1 \over 2}^+}\right
)(v')|\overline{c}\Gamma b|B(v)>\ \cong 2 \tau_{1/2}(w) \ {\rm Tr}
\left [ \overline{D}(v')\Gamma B(v)\right ]\nn \\ &&+\  {1 \over 2 m_b}
\Big \{ {\rm Tr}\left [ S_{\lambda}^{(b)}\overline{D}(v')\Gamma
\gamma^{\lambda}B(v)\right ] + \chi_{ke}^{(b)}\ {\rm Tr} \left [
\overline{D}(v')\Gamma B(v)\right ]\nn \\ &&+\  {\rm Tr}\left [
R_{\alpha\beta}^{(b)} \overline{D}(v')\Gamma P_+(v) i
\sigma^{\alpha\beta}B(v)\right ]\Big \} \eea

\noi where \bea \label{16e} &&D_{2^+}^{\sigma}(v') = P_+ (v')
\varepsilon_{v'}^{\sigma\nu} \gamma_{\nu} \nn \\ &&D_{1^+}^{\sigma}(v')
= - {\scriptstyle\sqrt{{3 \over 2}}}\ P_+ (v') \varepsilon_{v'}^{\nu}
\gamma_{5}
\left [ g_{\nu}^{\sigma} - {1 \over 3} \gamma_{\nu}\left (
\gamma^{\sigma} - v'^{\sigma}\right ) \right ] \nn \\ &&D_{1^+}(v') =
P_+ (v') \varepsilon_{v'}^{\nu} \gamma_5 \gamma_{\nu} \nn \\
&&D_{0^+}(v') = P_+ (v')\ . \eea

\noi The notations $S_{\sigma \lambda}^{(b)}$, $S_{\lambda}^{(b)}$
denote the perturbations to the current, and $\eta_{ke}^{(b)}$,
$\chi_{ke}^{(b)}$ and $R_{\sigma\alpha\beta}^{(b)}$ and
$R_{\alpha\beta}^{(b)}$ denote respectively the kinetic and the
magnetic perturbations to the Lagrangian. In the preceding relations
(\ref{15e}) $B(v)$ can be a pseudoscalar or a vector.\par

Expanded in terms of Lorentz covariant factors and subleading IW
functions, these tensor quantities read \cite{9r}~:
\bea \label{17e} &&S_{\sigma \lambda}^{(b)} = v_{\sigma} \left [
\tau_1^{(b)}(w) v_{\lambda} + \tau_2^{(b)}(w) v'_{\lambda} +
\tau_3^{(b)}(w) \gamma_{\lambda}\right ] + \tau_4^{(b)}(w)
g_{\sigma\lambda}\nn \\ &&S_{\lambda}^{(b)} = \zeta_1^{(b)}(w)
v_{\lambda} + \zeta_2^{(b)}(w) v'_{\lambda} + \zeta_3^{(b)}(w)
\gamma_{\lambda} \nn \\ &&R_{\sigma\alpha\beta}^{(b)} = \eta_1^{(b)}(w)
v_{\sigma}\gamma_{\alpha }\gamma_{\beta} + \eta_2^{(b)}(w)
v_{\sigma}v'_{\alpha }\gamma_{\beta } + \eta_3^{(b)}(w) g_{\sigma\alpha
}v'_{\beta }\nn \\ &&R_{\alpha\beta}^{(b)} = \chi_1^{(b)}(w)
\gamma_{\alpha }\gamma_{\beta } + \chi_2^{(b)}(w) v'_{\alpha
}\gamma_{\beta } \ . \eea

\noi The IW functions relevant to the current perturbation are not
independent, due to the equations of motion~:
\bea \label{18e} &&\tau_1^{(b)}(w) + w\tau_2^{(b)}(w) - \tau_3^{(b)}(w)
+ \tau_4^{(b)}(w) = 0 \nn \\ &&\zeta_1^{(b)}(w) + w\zeta_2^{(b)}(w) -
\zeta_3^{(b)}(w)  = 0 \eea

\noi and, at zero recoil, one has
\bea \label{19e} &&\tau_4^{(b)}(1) = \sqrt{3}\ \Delta E_{3/2}\
\tau_{3/2}(1) \nn \\ &&\zeta_3^{(b)}(1) =  - \Delta E_{1/2}\
\tau_{1/2}(1) \eea

\noi where a radial quantum number $n$ is implicit and $\Delta
E_{3/2}$, $\Delta E_{1/2}$ are the mass differences between the
excited states and the ground state. \par

Since, as pointed out above, we will also consider the intermediate
states $D\left ({3 \over 2}^-, 1^-\right )$, let us give the relevant
formulae, parallel to (\ref{15e})-(\ref{19e})
\bea \label{20e} &&<D\left ( {\scriptstyle {3 \over
2}^-}\right ) (v') |\overline{c}\Gamma b|B(v)> \ \cong \sqrt{3}\
\sigma_{3/2}(w)\ {\rm Tr} \left [
v_{\sigma}\overline{D}^{\sigma}(v')\Gamma B(v)\right ]\nn \\ &&+ \ {1
\over 2 m_b} \Big \{ {\rm Tr}\left [
T_{\sigma\lambda}^{(b)}\overline{D}^{\sigma}(v')\Gamma
\gamma^{\lambda}B(v)\right ] + \rho_{ke}^{(b)}\ {\rm Tr} \left [
v_{\sigma}\overline{D}^{\sigma}(v')\Gamma B(v)\right ]\nn \\ &&+ \ {\rm
Tr}\left [ V_{\sigma\alpha\beta}^{(b)} \overline{D}^{\sigma}(v')\Gamma
P_+(v) i \sigma^{\alpha\beta}B(v)\right ]\Big \} \nn \\
&&D_{1^-}^{\sigma} (v') = D_{1^+}^{\sigma} (v')(-
\gamma_5)\nn \\
&&T_{\sigma\lambda}^{(b)} = v_{\sigma}\left [ \sigma_1^{(b)}(w)
v_{\lambda} + \sigma_2^{(b)}(w) v'_{\lambda} + \sigma_3^{(b)}(w)
\gamma_{\lambda}\right ] + \sigma_4^{(b)}(w) g_{\sigma\lambda}\nn
\nn \\ &&V_{\sigma\alpha\beta}^{(b)} = \rho_1^{(b)}(w)
v_{\sigma}\gamma_{\alpha}\gamma_{\beta} + \rho_2^{(b)}(w)
v_{\sigma}v'_{\alpha}\gamma_{\beta} + \rho_3^{(b)}(w)
g_{\sigma\alpha}v'_{\beta}\nn \\   &&\sigma_1^{(b)}(w) +
w\sigma_2^{(b)}(w) - \sigma_3^{(b)}(w) + \sigma_4^{(b)}(w) = 0\nn \\
   &&\sigma_4^{(b)}(1) = \sqrt{3}\ \Delta E \left ( {\scriptstyle
{3 \over 2}^-}\right ) \sigma_{3/2}(1) \ . \eea

At zero recoil, Luke's theorem \cite{13r} imposes
\beq \label{21e} L_1(1) = L_2(1) = 0 \eeq

\noi while it can be shown that $L_4(1)$, $L_5(1)$, $L_6(1)$ are not
linearly independent \cite{8r}, and are related to two quantities,
namely
\beq \overline{\Lambda} = m_B - m_b = m_D - m_c \eeq

\noi  and the quantity called $\xi_3(1)$ by Falk and Neubert or
$\overline{\Sigma}$ by Uraltsev \cite{3r}~:
\bea \label{22e} &&L_4(1) = - \overline{\Lambda} + 2 \xi_3(1)\nn \\
&&L_5(1) = - \overline{\Lambda} \nn \\ &&L_6(1) = - \overline{\Lambda}
- \xi_3(1) \ . \eea

We demonstrate the identity $\xi_3(1) \equiv \overline{\Sigma}$ in
Appendix A. Considering the forward amplitude, i.e. taking $w_{if} =
1$, two SR can be obtained for
subleading corrections at zero recoil, as we will see below~:
\bea \label{23e} &&\overline{\Lambda} = 2 \sum_n \Delta
E_{1/2}^{(n)}|\tau_{1/2}^{(n)}(1)|^2 + 4 \sum_n \Delta
E_{3/2}^{(n)}|\tau_{3/2}^{(n)}(1)|^2\nn \\ &&\xi_3(1) \equiv
\overline{\Sigma} = 2 \sum_n \Delta
E_{3/2}^{(n)}|\tau_{3/2}^{(n)}(1)|^2 - 2 \sum_n \Delta
E_{1/2}^{(n)}|\tau_{1/2}^{(n)}(1)|^2 \eea

\noi where $\Delta E_j^{(n)}$ and $\tau_j^{(n)}(1)$ ($j = {1 \over 2},
{3 \over 2})$ are the corresponding level spacings and transition IW
functions between the ground state ${1\over 2}^-$ and the $P$-wave
states ${1\over 2}^+$ and ${3\over 2}^+$. The first SR is Voloshin SR
\cite{14r}, and the second one was discovered by A. Le Yaouanc et al.
\cite{12r} and by Uraltsev \cite{3r}. We have adopted the notation of
Isgur and Wise for the transition IW functions \cite{2r}.

\section{B Meson sum rule.} \hspace*{\parindent} We take as initial and
final states the ground state pseudoscalar meson at different
four-velocities $B(v_i)$ and $B(v_f)$ and, as in \cite{3r}-\cite{6r},
the axial currents aligned along the corresponding four-velocities,
$\Gamma_i = {/\hskip - 2
truemm v}_i\gamma_5$ and $\Gamma_f = {/\hskip - 2 truemm
v}_f\gamma_5$. Then, the subleading SR (\ref{11e}) writes~:
\bea \label{28e} &&\sum_n \Big \{ (w_iw_f-w_{if})  \xi^{(n)}(w_i) \left [
L_1^{(n)}(w_f)+L_4^{(n)}(w_f)\right ] \nn \\
&&+ \ (1-w_i) 2 \tau_{1/2}^{(n)}(w_i) F_{1/2}^{(n)}(w_f) \nn \\
&&+ \left [ (w_iw_f-w_{if})^2 - {1 \over 3} (w_i^2 - 1) (w_f^2 - 1)
\right ] \sqrt{3}\ \tau_{3/2}^{(n)}(w_i) \ F_{3/2}^{(n)}(w_f) \nn \\
&&+ \ (w_iw_f-w_{if}) \sqrt{3}\ \sigma_{3/2}^{(n)}(w_i) \
G_{3/2}^{(n)}(w_f) + \hbox{Higher $j^P$ states} + (i \leftrightarrow
f) \Big \}\nn \\
&&= - \left [ 2L_1(w_{if})(1 +
w_{if} - w_i -w_f)-2L_4(w_{if})(1 - w_{if})\right ] \eea

The first, second, third and fourth term in the l.h.s. of (\ref{28e})
correspond to the $\left ( {1\over 2}^-, 1^- \right )$, $\left (
{1\over 2}^+, 0^+ \right )$, $\left ( {3\over 2}^+, 2^+ \right )$ and
$\left ( {3\over 2}^-, 1^- \right )$ intermediate $D$ states. No other
states $j^P$ with $j \leq {3 \over 2}$ appear in the l.h.s. because of
the number of $\gamma_5$ matrices involved in the traces over Dirac
matrices.\par

In equation (\ref{28e}) we have made explicit the subleading ${1\over
2}^- \to {1 \over 2}^-$ elastic functions $L_j^{(n)}$ $(j=1,4)$ and we
have factorized, when possible, for the contributions ${1\over 2}^- \to
{1 \over 2}^+, {3\over 2}^+$ and ${3\over 2}^-$, polynomials in
$(w_i,w_f,w_{if})$ that vanish at the frontier $(w,1,w)$ of the domain
(\ref{5e}). The IW functions $\tau_{1/2}^{(n)}(w)$, $\tau_{3/2}^{(n)}(w)$ and
$\sigma_{3/2}^{(n)}(w)$ do not
vanish at $w=1$. Similarly, the complicated functions $F_{1/2}^{(n)}(w)$,
$F_{3/2}^{(n)}(w)$ and $G_{3/2}^{(n)}(w)$ are given in terms of the
form factors
defined in Section 3 and do not vanish in general for $w=1$. \par

It is not necessary to give all the explicit expressions of these functions
since we are interested in the frontier of the domain $(w_i,w_f,w_{if})
= (w,1,w)$ and, as we will see below, only $F_{1/2}^{(n)}(1)$ will
contribute.\par

There are two crucial features in expression (\ref{28e}). First, the
appearance of the subleading functions $L_1(w_{if})$, $L_4(w_{if})$ in
the r.h.s., since we consider the whole allowed domain for the
variables $(w_i, w_f, w_{if})$. Second, the polynomials in
$(w_i,w_f,w_{if})$, that result from the sum over the spin $J = 1,2$
polarizations \cite{4r}~:
\bea \label{31e} &&\sum_{\lambda} \varepsilon_{v'}^{(\lambda)*\mu}\
\varepsilon_{v'}^{(\lambda)\nu}\ v_{f{\mu}} v_{i{\nu }} = w_i w_f -
w_{if} \nn \\ &&\sum_{\lambda} \varepsilon_{v'}^{(\lambda)*\mu\nu }\
\varepsilon_{v'}^{(\lambda)\rho\sigma}\ v_{f{\mu}} v_{f{\nu }}
v_{i{\rho }} v_{i{\sigma }}= (w_i w_f - w_{if})^2 - {1 \over 3}
(w_i^2-1)(w_f^2-1) \eea

\noi where $\lambda$ runs over the $2J+1$ polarizations.\par

These polynomials will imply the vanishing of the corresponding
contributions at $(w_i, w_f, w_{if}) = (w, 1,w)$. This will occur
also for higher $j^P$ intermediate
states, because one obtains, in all generality \cite{4r}, for the
projector on the polarization tensor of a particle of integer spin $J$,
contracted with $v_i$ and $v_f$ four-velocities~:
\beq
\label{equation44}
T^{\nu_1 \cdots \nu_J, \mu_1 \cdots \mu_J}_{v'} = \sum_{\lambda}
\varepsilon_{v'}^{(\lambda )*\nu_1 \cdots \nu_J} \
\varepsilon_{v'}^{(\lambda )\mu_1 \cdots \mu_J}
\eeq
\bea \label{34e}
&&v_{f\nu_1} \cdots v_{f \nu_J}  \ T_{v'}^{\nu_1 \cdots \nu_J, \mu_1
\cdots \mu_J} \ v_{i\mu_1}
\cdots v_{i \mu_J} = \nn \\ &&\sum_{k=0}^{J/2} (-1)^k {(J!)^2 \over
(2J)!}\ {(2J - 2k)! \over k!(J-k)!(J-2k)!} (w_i^2 - 1)^k (w_f^2 - 1)^k
(w_iw_f - w_{if})^{J-2k} \eea

\noi that vanishes for $J > 0$ at $(w_i,w_f, w_{if}) = (w, 1, w)$. \par

Therefore, at the frontier
\beq
(w_i,w_f,w_{if}) \to (w,1,w)
\eeq

\noi the SR (\ref{28e}) will write, very simply, dividing by a factor $(w-1)$
\beq
L_4(w) = \sum_n \tau_{1/2}^{(n)} (w) F_{1/2}^{(n)}(1)\ .
\eeq

We only need the functions $F_{1/2}^{(n)}(w)$ for $w=1$.
The calculation gives, for all $w$,
\bea
F_{1/2}^{(n)}(w) &=& (1- w) \Big [ \zeta_1^{(b)(n)}(w) + (1 + 2w)
\zeta_2^{(b)(n)}(w) + \chi_{kin}^{(b)(n)}(w) + 6 \chi_1^{(b)(n)}(w)
\nn \\
&&- 2(1+w)  \chi_2^{(b)(n)}(w)\Big ] + 2(1+2w) \zeta_3^{(b)(n)}(w)
\eea

\noi and for $w= 1$,
\beq
F_{1/2}^{(n)}(1) = 6 \zeta_3^{(b)(n)}(1)
\eeq

\noi and from the relation (\ref{19e}) we obtain finally
\beq
L_4(w) = - 6 \sum_n \Delta E_{1/2}^{(n)}\ \tau_{1/2}^{(n)}(1)\
\tau_{1/2}^{(n)}(w)\ .
\eeq

\section{B$^{\bf *}$ meson sum rule.} \hspace*{\parindent} We take as
initial and final states the ground state vector meson at different
four-velocities $B^{*(\lambda_i)}(v_i)$ and $B^{*(\lambda_f)}(v_f)$ and
we adopt the particular case of the $B^*$ polarizations (see appendix A of
ref. \cite{4r})~:
\beq \label{36e} \varepsilon_i = {v_f - w_{if}v_i \over \sqrt{w_{if}^2
- 1}} \qquad \qquad \varepsilon_f = {v_i - w_{if}v_f \over
\sqrt{w_{if}^2 - 1}} \eeq

\noi that satisfy $\varepsilon_i \cdot v_i = \varepsilon_f\cdot v_f =
0$ and $\varepsilon_i^2 = \varepsilon_f^2 = -1$. With the definitions
(\ref{36e}) one has
$\varepsilon_i\cdot \varepsilon_f = w_{if}$, but we can change one
global sign in (\ref{36e}) to make $\varepsilon_i\cdot \varepsilon_f
= - w_{if}$ and
therefore $\varepsilon_i \cdot \varepsilon_f \to - 1$ when $v_i \to
v_f$. The sum rules, being linear in $\varepsilon_i$ and in
$\varepsilon_f$, do not depend on this overall sign. \par

Then, performing the relevant traces, the subleading SR (\ref{11e}) writes
\bea \label{37e} &&\sum_n \Big \{\left ( \varepsilon_f^*\cdot
v'\right ) \left (
\varepsilon_i\cdot v'\right )  \xi^{(n)}
(w_i) \left [ L_2^{(n)}(w_f)-L_5^{(n)}(w_f) \right . \nn \\
&&\left . - (1- w_f) L_3^{(n)}(w_f) + (1 +
w_f) L_6^{(n)}(w_f) \right ] \nn \\
&&+\  2 \tau_{1/2}^{(n)}(w_i) \left [ \left ( \varepsilon_f^*\cdot
\varepsilon_i\right )
K_1^{(n)}(w_f, w_i)  + \left ( \varepsilon_i\cdot v_f\right )\left (
\varepsilon_f^*\cdot v'\right )
K_2^{(n)}(w_f,w_i) \right .\nn \\ &&\left . + \left (
\varepsilon_f^*\cdot v_i\right ) \left ( \varepsilon_i\cdot
v'\right ) K_3^{(n)} (w_f)
    + \left ( \varepsilon_i\cdot v'\right )\left (
\varepsilon_f^*\cdot v'\right ) K_4^{(n)}(w_f,w_i,w_{if}) \right ] \nn \\
&&+  \ \sqrt{3}\ \tau_{3/2}(w_i) \left [ \left (
\varepsilon_f^*\cdot \varepsilon_i\right ) S_1^{(n)}(w_f,w_i)  + \left (
\varepsilon_f^*\cdot v'\right )\left ( \varepsilon_i\cdot v_f\right )
S_2^{(n)}(w_f,w_i)
\right .\nn \\ &&\left . + \left ( \varepsilon_f^*\cdot
v_i\right ) \left ( \varepsilon_i\cdot v'\right )S_3^{(n)} (w_f,w_i)  + \left
( \varepsilon_i\cdot v'\right )\left ( \varepsilon_f^*\cdot v'\right
) S_4^{(n)}(w_f,w_i,w_{if}) \right ] + (i \leftrightarrow f) \Big \}\nn \\
&&+ \ \sqrt{3} \ \tau_{3/2}^{(n)}(w_i) T^{(n)}(w_f) \Big [
(w_{if}-w_iw_f)Tr \left [ \gamma^{\mu} {/ \hskip - 2 truemm v}' {/
\hskip - 2 truemm v}_f{/ \hskip - 2 truemm \varepsilon}_f^*
\gamma_5\right ] Tr \left [ \gamma_{\mu} {/ \hskip - 2 truemm v}' {/
\hskip - 2 truemm v}_i {/ \hskip - 2 truemm \varepsilon}_i \gamma_5
\right ] \nn \\
&&+\ Tr \left [ {/ \hskip - 2 truemm v}_f {/ \hskip - 2 truemm v}' {/
\hskip - 2 truemm v}_i {/ \hskip - 2 truemm \varepsilon}_i
\gamma_5\right ] Tr \left [ {/ \hskip - 2 truemm v}_i {/ \hskip - 2
truemm v}' {/ \hskip - 2 truemm v}_f {/ \hskip - 2 truemm
\varepsilon}_f^*\gamma_5 \right ] \Big ]  \nn \\
&&+ \ \sqrt{3} \sigma_{3/2}^{(n)}(w_i) U^{(n)}(w_f, w_i) \ Tr \left [
\gamma^{\mu} {/\hskip - 2 truemm v}'
{/\hskip - 2 truemm v}_f  {/ \hskip - 2 truemm \varepsilon}_f^*
\gamma_5 \right ] \ Tr \left [ \gamma_{\mu} {/\hskip - 2 truemm v}'
{/\hskip - 2 truemm v}_i {/ \hskip - 2 truemm \varepsilon}_i
\gamma_5 \right ] \nn \\
&&+\  \sqrt{3} \sigma_{3/2}^{(n)}(w_i) \left [ v_{f \mu}
\varepsilon^*_{f \nu} T_{v'}^{\mu\nu, \rho \sigma} v_{i \rho}
\varepsilon_{i\sigma} V_1^{(n)}(w_f, w_i)\right . \nn \\
&&+ \ v_{f \mu} \varepsilon_{f\nu}^* T_{v'}^{\mu \nu , \rho\sigma}
v_{i\rho} v_{i\sigma} (\varepsilon_i \cdot v') V_2^{(n)}(w_f, w_i)\nn
\\
&&+ \ v_{f \mu} v_{f \nu} T_{v'}^{\mu \nu , \rho\sigma} v_{i\rho}
\varepsilon_{i \sigma} (\varepsilon_f^* \cdot v')
V_3^{(n)}(w_f,w_i)\nn \\
&&\left . + \ v_{f \mu} v_{f\nu} T_{v'}^{\mu \nu , \rho\sigma}
v_{i\rho} v_{i\sigma} (\varepsilon_f^* \cdot v') (\varepsilon_i \cdot
v') V_4^{(n)}(w_f,w_i)\right ]\nn \\
&&+ \ \hbox{Higher $j^P$ intermediate states} + \ (i \leftrightarrow
f)  \Big \}\nn \\
&&= - \Big \{ - 2\left [ \left (
\varepsilon_f^*\cdot \varepsilon_i\right ) (1 - w_i - w_f +
w_{if}) + \left ( \varepsilon_f^*\cdot v'\right
) \left ( \varepsilon_i\cdot v_f\right )\right . \nn \\
&&\left . + \left ( \varepsilon_f^*\cdot
v_i\right ) \left ( \varepsilon_i\cdot v'\right ) - \left (
\varepsilon_f^*\cdot v_i\right ) \left ( \varepsilon_i\cdot v_f\right
)\right ] L_2(w_{if}) \nn \\
&&- \left [ \left ( \varepsilon_i\cdot
v_f\right ) \left ( \varepsilon_f^*\cdot v'\right ) (w_{if}- 1) + \left
( \varepsilon_f^*\cdot v_i\right ) \left ( \varepsilon_i\cdot v'\right
) (w_{if}-1) \right . \nn \\
&&\left . - \left ( \varepsilon_i\cdot
v_f\right ) \left ( \varepsilon_f^*\cdot v_i\right ) (w_i + w_f - 2)
\right ] L_3(w_{if})  \nn \\
&&+ \ 2 \left [ \left ( \varepsilon_f^*\cdot
\varepsilon_i\right ) (1 - w_{if}) +  \left ( \varepsilon_f^*\cdot
v_i\right ) \left ( \varepsilon_i\cdot v_f\right )\right ] L_5(w_{if})  \nn \\
&&+\left [ - \left ( \varepsilon_i\cdot v_f\right ) \left (
\varepsilon_f^*\cdot v'\right ) (w_{if} + 1) - \left (
\varepsilon_f^*\cdot v_i\right ) \left ( \varepsilon_i\cdot v'\right )
(w_{if} + 1) \right . \nn \\
&&\left . + \left ( \varepsilon_i\cdot
v_f\right ) \left ( \varepsilon_f^*\cdot v_i\right )(w_i + w_f
-2)\right ]  L_6(w_{if})\Big \}\eea

In the l.h.s. of equation (\ref{37e}), the first term corresponds to the
intermediate states $\left ( {1 \over 2}^-, 0^-\right ) + \left ( {1 \over
2}^-, 1^-\right )$ (both spins contribute due to the fact that one has
more four-vectors and a $\gamma_5$ in the traces than in the
pseudoscalar case), the second, third and fourth terms correspond to the
contributions $\left ( {1 \over 2}^+, 1^+\right )$, $\left ( {3
\over 2}^+, 1^+\right )$ and $\left ( {3 \over 2}^+, 2^+\right )$ and
the fifth and sixth terms correspond to the contributions $\left ( {3
\over 2}^-, 1^-\right )$ and $\left ( {3 \over 2}^-, 2^-\right )$. We
have made explicit the subleading ${1\over 2}^-
\to {1 \over 2}^-$ functions $L_j^{(n)}$ $(j=2,3,5,6)$ and we
have kept the explicit dependence on the initial and final polarizations
$\varepsilon_i$, $\varepsilon_f$. This allows to factorize, when
possible, for the contributions ${1\over 2}^- \to {1 \over 2}^+$,
${3\over 2}^+$ and  ${3 \over 2}^-$, polynomials in $(w_i,w_f,w_{if})$
that vanish at the frontier $(w, 1,w)$. The
IW functions $\tau_{1/2}^{(n)}(w)$, $\tau_{3/2}^{(n)}(w)$ and
$\sigma_{3/2}^{(n)}(w)$
   that do not vanish in
general for $w=1$, and the complicated functions $K_j^{(n)}$ $(j=1,\cdots
4)$, $S_j^{(n)}$ $(j=1,\cdots 4)$, $T^{(n)}$,
$U^{(n)}$ and $V_j^{(n)}$ $(j=1,\cdots 4)$ can be computed in terms
of the form factors
defined in Section 3. \par

The tensor $T^{\mu \nu , \rho \sigma}_{v'}$, that appears in the
l.h.s. of eq. (\ref{37e}), is given in terms of the $J=2$
polarization tensors by the expression
\beq
T_{v'}^{\mu \nu , \rho \sigma} = \sum_{\lambda}
\varepsilon_{v'}^{(\lambda ) \mu \nu} \ \varepsilon_{v'}^{(\lambda )
\rho \sigma}
\eeq

In order to see clearly which terms survive at the frontier
$(w_i,w_f,w_{if}) = (w,1,w)$, it is not necessary to go to the details
that we have given in Section 4 for the sake of clarity. It is enough
to realize that the limit
\beq
(w_i,w_f,w_{if}) \to (w,1,w)
\eeq

\noi corresponds to the limit
\beq
\label{equation55}
   v_f \to v'  \quad , \qquad v_i \to v
\eeq

\noi $(v\cdot v' = w)$ and make use of the orthogonality conditions between the
intermediate states polarization tensors and $v'$.
Explicit calculations confirm this simple argument. \par

   In the limit (\ref{equation55}), we have
\bea
\label{equation56}
&&\varepsilon_f^* \cdot \varepsilon_i \to w \nn \\
&&\varepsilon_f^* \cdot v' \to 0 \nn \\
&&\varepsilon_i \cdot v' \to - \sqrt{w^2-1} \nn \\
&&\varepsilon_f^* \cdot v_i \to - \sqrt{w^2-1} \nn \\
&&\varepsilon_i \cdot v_f \to - \sqrt{w^2-1} \nn \\
&&Tr \left [  \gamma^{\mu}  {/\hskip - 2 truemm v}'
{/\hskip - 2 truemm v}_f  {/ \hskip - 2 truemm \varepsilon}_f^*
\gamma_5 \right ] \to 0\nn \\
&&Tr \left [ {/\hskip - 2 truemm v}_f  {/\hskip - 2 truemm v}'
{/\hskip - 2 truemm v}_i  {/ \hskip - 2 truemm \varepsilon}_i
\gamma_5 \right ] , Tr \left [ {/\hskip - 2 truemm v}_i {/\hskip - 2 truemm v}'
{/\hskip - 2 truemm v}_f  {/ \hskip - 2 truemm \varepsilon}_f^*
\gamma_5 \right ]\to 0\nn \\
&&T_{v'}^{\mu \nu , \rho \sigma} \ v_{f\mu} \varepsilon_{f\nu}^*
v_{i\rho} \varepsilon_{i\sigma} \to 0 \nn \\
&&T_{v'}^{\mu \nu , \rho \sigma} \ v_{f\mu} v_{f\nu} v_{i\rho}
\varepsilon_{i\sigma} \to 0 \nn \\
&&T_{v'}^{\mu \nu , \rho \sigma} \ v_{f\mu} \varepsilon_{f\nu}^*
v_{i\rho} v_{i\sigma} \to 0 \nn \\
&&T_{v'}^{\mu \nu , \rho \sigma} \ v_{f\mu} v_{f\nu} v_{i\rho}
v_{i\sigma} \to 0 \ .
\eea

\noi The last limits follow from the orthogonality condition $T^{\mu \nu ,
\rho \sigma} v'_{\rho} = T^{\mu \nu , \rho \sigma} v'_{\mu} = 0$.
Notice that the ``Higher $j^P$ intermediate states'' contributions in
equation (\ref{37e}), similarly to the four last expressions (\ref{equation56})
and to the general expression (\ref{34e}), due to the symmetry in
($\nu_1,\nu_2, \cdots \nu_J$) and in ($\mu_1,\mu_2 ,\cdots \mu_J$) and the
linearity in $\varepsilon_i$ and $\varepsilon_f^*$, will be proportional
to the following quantities
\bea
\label{equation57}
&&T_{v'}^{\nu_1\nu_2\cdots \nu_J, \mu_1\mu_2\cdots \mu_J}\ v_{f\nu_1}
v_{f\nu_2} \cdots \varepsilon_{f\nu_J}^*\ v_{i\mu_1} v_{i\mu_2}
\cdots v_{i\mu_{J-1}} \varepsilon_{i\mu_J} \nn \\
&&T_{v'}^{\nu_1\nu_2\cdots \nu_J, \mu_1\mu_2\cdots \mu_J}\ v_{f\nu_1}
v_{f\nu_2} \cdots v_{f\nu_J}\ v_{i\mu_1} v_{i\mu_2} \cdots
v_{i\mu_{J-1}} \varepsilon_{i\mu_J} \nn \\
&&T_{v'}^{\nu_1\nu_2\cdots \nu_J, \mu_1\mu_2\cdots \mu_J}\ v_{f\nu_1}
v_{f\nu_2} \cdots \varepsilon_{f\nu_J}^*\ v_{i\mu_1} v_{i\mu_2}
\cdots v_{i\mu_{J-1}} v_{i\mu_J} \nn \\
&&T_{v'}^{\nu_1\nu_2\cdots \nu_J, \mu_1\mu_2\cdots \mu_J}\ v_{f\nu_1}
v_{f\nu_2} \cdots v_{f\nu_J}\ v_{i\mu_1} v_{i\mu_2} \cdots 
v_{i\mu_{J-1}} v_{i\mu_J}\ . \eea

\noi The tensor $T_{v'}^{\nu_1\nu_2
\cdots \nu_J, \mu_1\mu_2 \cdots \mu_J}$ is given, in terms of
the polarization tensor of an intermediate state of spin $J$, by
expression (\ref{equation44}), and the last quantity in
(\ref{equation57}) is given by the
polynomial in $(w_i,w_f,w_{if})$ (\ref{34e}).\par

By the same argument as before, due to the orthogonality conditions
\beq
T_{v'}^{\nu_1\nu_2\cdots \nu_J, \mu_1\mu_2\cdots \mu_J}\ v'_{\mu_k} =
T_{v'}^{\nu_1\nu_2\cdots \nu_J, \mu_1\mu_2\cdots \mu_J}\ v'_{\nu_k} =
0 \qquad (k = 1\cdots J)
\eeq

\noi all the quantities (\ref{equation57}) go to $0$ in the limit
$v_f \to v'$ (\ref{equation55}). \par

Therefore, in the limit $(w_i,w_f,w_{if}) \to (w,1,w)$, the SR
(\ref{37e}), taking into account its symmetry in $i \leftrightarrow f$ and
the {\it asymmetry} in $i \leftrightarrow f$ of the limit
(\ref{equation55}), becomes
the much simpler expression~:
\bea
\label{equation59}
&&\sum_n \Big \{  2 \tau_{1/2}^{(n)}(w) \left [ wK_1^{(n)}(1,w) +
(w^2 - 1) K_3^{(n)} (1)\right ] \nn \\
&&+ \ 2 \tau_{1/2}^{(n)}(1) \left [ wK_1^{(n)}(w,1) + (w^2 - 1)
K_2^{(n)} (w, 1)\right ]\Big \} \nn \\
&&+ \ \Big \{ \sqrt{3}\  \tau_{3/2}^{(n)}(w) \left [ wS_1^{(n)}(1,w)
+ (w^2 - 1) S_3^{(n)} (1, w)\right ] \nn \\
&&+ \ \sqrt{3}\  \tau_{3/2}^{(n)}(w)\left [ wS_1^{(n)}(w,1) + (w^2 -
1) S_2^{(n)} (w, 1)\right ]\Big \} \nn \\
&&= - 2 (w-1) L_5(w) + 2(w^2 - 1) L_6(w) \ .
\eea

Therefore, we have only to compute the functions
$K_1^{(n)}(w_i,w_f)$, $K_2^{(n)}(w_i,w_f)$, $K_3^{(n)}(w_i)$,
$S_1^{(n)}(w_i,w_f)$, $S_2^{(n)}(w_i,w_f)$ and $S_3^{(n)}(w_i,w_f)$.
The explicit calculation gives, for general $(w_i,w_f)$,
\bea
&&K_1^{(n)}(w_i,w_f) = - (w_i - 1) (w_f - 1) \Big [
\zeta_1^{(b)(n)}(w_i) + (2w_i+1) \zeta_2^{(b)(n)}(w_i) \nn \\
&&+ \chi_{kin}^{(b)(n)}(w_i) - 2\chi_1^{(b)(n)}(w_i)\Big ] - 2(1-w_f)
w_i\zeta_3^{(b)(n)}(w_i) \nn \\
&&K_2^{(n)}(w_i,w_f) = - (1- w_f) \Big [ \zeta_1^{(b)(n)}(w_i) +
(2w_i+1) \zeta_2^{(b)(n)}(w_i) - 2\zeta_3^{(b)(n)}(w_i)\nn \\
&&+ \chi_{kin}^{(b)(n)}(w_i) - 2\chi_1^{(b)(n)}(w_i) +
2\chi_2^{(b)(n)}(w_i)\Big ] \nn \\
&&K_3^{(n)}(w_i) = - (1- w_i) \Big [ \zeta_1^{(b)(n)}(w_i) + (2w_i+1)
\zeta_2^{(b)(n)}(w_i) \nn \\
&&+ \chi_{kin}^{(b)(n)}(w_i) - 2\chi_1^{(b)(n)}(w_i)\Big ] - 2
w_i\zeta_3^{(b)(n)}(w_i)\eea

\noi and
\bea
&&S_1^{(n)}(w_i,w_f) = - {1 \over 6} (w_f^2 - 1) G_1^{(n)}(w_i)\nn \\
&&S_2^{(n)}(w_i,w_f) = - {1 \over 6} (w_f^2 - 1) G_2^{(n)}(w_i)\nn \\
&&S_3^{(n)}(w_i,w_f) = - {1 \over 6} (2 - w_f) G_1^{(n)}(w_i)
\eea

\noi where in the preceding equations
\bea
&&G_1^{(n)}(w_i) = (w_i^2 - 1) \Big [ \tau_{1}^{(b)(n)}(w_i) +
(2w_i+1) \tau_{2}^{(b)(n)}(w_i) - 2\tau_{3}^{(b)(n)}(w_i)  \nn \\
&&+ \chi_{kin}^{(b)(n)}(w_i) - 2 \eta_{1}^{(b)(n)}(w_i) -
3\eta_{3}^{(b)(n)}(w_i)\Big ] + 2(w_i+1)^2 \tau_{4}^{(b)(n)}(w_i)\nn
\\
&& \nn \\
&&G_2^{(n)}(w_i) = (2 - w_i) \tau_{1}^{(b)(n)}(w_i) + (2w_i+1)
(2-w_i)  \tau_{2}^{(b)(n)}(w_i) +2w_i \tau_{3}^{(b)(n)}(w_i)  \nn \\
&&-2w_i \tau_{4}^{(b)(n)}(w_i) + (2-w_i)  \chi_{kin}^{(b)(n)}(w_i) -
2 (2-w_i) \eta_{1}^{(b)(n)}(w_i) \nn \\
&&+ 4(w_i -1) \eta_{2}^{(b)(n)}(w_i) + 3(w_i -2) \eta_3^{(b)(n)}(w_i)
\eea

\noi and we get, for the quantities needed in equation (\ref{equation59}),
\bea
&&K_1^{(n)}(1,w) = - 2(1-w) \zeta_3^{(b)(n)}(1) \nn \\
&&K_3^{(n)}(1) = - 2\zeta_3^{(b)(n)}(1) \nn \\
&&K_1^{(n)}(w,1) = K_2^{(n)}(w,1) = 0 \nn \\
&&S_1^{(n)}(1,w) = - {4 \over 3} (w^2-1) \tau_4^{(b)(n)}(1) \nn \\
&&S_3^{(n)}(1,w) = - {4 \over 3} (2-w) \tau_4^{(b)(n)}(1) \nn \\
&&S_1^{(n)}(w,1) = S_2^{(n)}(w,1) = 0
\eea

\noi that gives, dividing the SR by the factor $2(w-1)$,
\bea
&&-L_5(w) + (w+1)L_6(w)\nn \\
&&= - \sum_n  \left [ 2\tau_{1/2}^{(n)}(w)\ \zeta_3^{(b)(n)}(1) + {4
\over 3} (w+1) \sqrt{3}\  \tau_{3/2}^{(n)}(w)\
\tau_4^{(b)(n)}(1)\right ]
\eea

\noi and using (\ref{19e}) one gets finally~:
\bea
&&-L_5(w) + (w+1)L_6(w)\nn \\
&&=  2 \sum_n \Delta E_{1/2}^{(n)}\ \tau_{1/2}^{(n)}(w)\
\tau_{1/2}^{(n)}(1) - 4(w+1)\sum_n \Delta E_{3/2}^{(n)}\
\tau_{3/2}^{(n)}(w) \ \tau_{3/2}^{(n)}(1) \ .
\eea

\section{Basic results.} \hspace*{\parindent}
Let us recall the two sum rules that we have obtained in the two
preceding sections~:
\beq \label{33e} L_4(w) = - 6 \sum_n \Delta E_{1/2}^{(n)} \
\tau_{1/2}^{(n)}(1) \ \tau_{1/2}^{(n)}(w) \eeq
\bea \label{42e} - L_5(w) + (w+1) L_6(w) &=& 2 \sum_n \Delta E_{1/2}^{(n)} \
\tau_{1/2}^{(n)}(1) \ \tau_{1/2}^{(n)}(w)\nn \\ &&- \ 4(w+1)  \sum_n
\Delta E_{3/2}^{(n)} \ \tau_{1/2}^{(n)}(1) \ \tau_{3/2}^{(n)}(w) \ . \eea

Due to the equations of motion, the functions $L_i(w)$ $(i = 4,5,6)$
are not independent and, as shown
in ref. \cite{8r}, are given in terms of the
elastic IW function $\xi (w)$, a subleading function $\xi_3(w)$ and
the
$\overline{\Lambda}$ parameter ($\overline{\Lambda} = m_B - m_b$)~:
\bea \label{43e} &&L_4(w) = - \overline{\Lambda} \xi (w) + 2 \xi_3(w)
\nn \\ &&L_5(w) = - \overline{\Lambda} \xi (w) \nn \\ &&L_6(w) = - {2
\over w + 1} \left [ \overline{\Lambda} \xi (w) + \xi_3(w) \right ] \ .
\eea

\noi Therefore, from (\ref{33e})-(\ref{43e}) we obtain the
interesting relations, {\it valid for all w}~:
\bea \label{44e} \overline{\Lambda} \xi (w) &=& 2(w+1) \sum_n \Delta
E_{3/2}^{(n)} \ \tau_{3/2}^{(n)}(1) \ \tau_{3/2}^{(n)}(w)\nn \\ &&+ \ 2
\sum_n \Delta E_{1/2}^{(n)} \ \tau_{1/2}^{(n)}(1) \ \tau_{1/2}^{(n)}(w)
\eea
\bea \label{45e} \xi_3 (w) &=& (w+1) \sum_n \Delta E_{3/2}^{(n)} \
\tau_{3/2}^{(n)}(1) \ \tau_{3/2}^{(n)}(w)\nn \\ &&- \ 2 \sum_n \Delta
E_{1/2}^{(n)} \ \tau_{1/2}^{(n)}(1) \ \tau_{1/2}^{(n)}(w) \ . \eea

These remarkably simple relations are the basic results of the present
paper. They reduce to the known results (\ref{23e}) for $w=1$. It is 
important to notice that both the {\it subleading}
quantities $\overline{\Lambda}\xi (w)$ and $\xi_3(w)$ can be
expressed in terms of {\it leading} quantities, namely the IW
functions $\tau_j^{(n)}(w)$ and
the level spacings $\Delta E_j^{(n)}$ ($j = {1 \over 2}, {3 \over 2})$.
\par

Very remarkably, equation (\ref{44e}) shows that the leading IW
function $\xi (w)$ appears constrained
to be a combination of the {\it averages}
\beq \label{46e} {1 \over \overline{\Lambda}} \sum_n \Delta E_j^{(n)}
\tau_j^{(n)}(1) \tau_j^{(n)} (w) \qquad \left ( j = {\scriptstyle{1
\over 2}}, {\scriptstyle{3 \over 2}}\right )
\eeq

\noi or, conversely, the fundamental constant $\overline{\Lambda}$ is 
given by the ratio of {\it
functions}~:
\bea \label{47e} \overline{\Lambda} &=& {1 \over \xi (w)} \left [
2(w+1) \sum_n \Delta E_{3/2}^{(n)} \ \tau_{3/2}^{(n)}(1) \
\tau_{3/2}^{(n)}(w)\right .\nn \\ &&\left . + \ 2  \sum_n \Delta
E_{1/2}^{(n)} \ \tau_{1/2}^{(n)}(1) \ \tau_{1/2}^{(n)}(w)\right ] \ .
\eea

It is worth to underline that all other subleading IW functions
except $\zeta_3^{(b)}(1)$ cannot contribute
to the l.h.s. of the SR (\ref{28e}) because the polynomials
(\ref{31e}) vanish at the
frontier of the domain $(w_i, w_f, w_{if}) = (w, 1, w)$. In the case
of the $B^*$ SR, equation
(\ref{37e}), for the same reason only $\zeta_3^{(b)}(1)$ and
$\tau_4^{(b)}(1)$ survive in the l.h.s. at $(w_i,w_f,w_{if}) = (w, 1,
w)$.

\section{Phenomenological discussion.} \hspace*{\parindent} To
illustrate our results for the subleading form factors, we will
concentrate on some functions that play a role in the analysis of $B
\to D(D^*)\ell\nu$, and about which we can get information from the
relations obtained in the preceding Section on $\overline{\Lambda}\xi
(w)$ and $\xi_3(w)$. Of particular interest are the functions
\beq \label{eqnew} L_4(w) \qquad \hbox{and} \qquad \eta (w) = {\xi_3(w)
\over \overline{\Lambda}\xi (w)} \eeq

\noi and their values and derivatives at zero recoil. \par

The function $L_4(w)$ appears at first order in $1/m_Q$ in the
differential semi-leptonic rate of $B \to D\ell \nu$ \cite{8r}. This
subleading IW function is specially important, but can be expressed,
from (\ref{43e}), in terms of $\xi (w)$, $\overline{\Lambda}$ and the
commonly used function $\eta (w)$ (see for example \cite{11r} and
references therein)~:
\beq
L_4(w) = - \overline{\Lambda} \xi (w) \left [ 1 - 2 \eta (w) \right ]
\eeq

\subsection{Check of the Bakamjian-Thomas quark models.}
\hspace*{\parindent}
We would like first to test whether the results found in the class
of relativistic quark models of the Bakamjian-Thomas type \cite{15r}
are consistent with the sum rules found in this paper, in particular the $w$
dependence.
This class of models yield covariant form factors in the heavy quark
limit that satisfy Isgur-Wise scaling, and Bjorken and Uraltsev SR. It
is a class of models in the sense that one can choose the dynamics in
the hadron rest frame, and then compute the corresponding Isgur-Wise
functions with the boosted wave functions.\par

The dynamics at rest that describes in the most accurate way the
$Q\overline{Q}$, $Q\overline{q}$ and $q\overline{q}$ spectra (where $Q$
and $q$ denote respectively heavy and light quarks) is the
phenomenological Hamiltonian set up by Godfrey and Isgur \cite{16r},
containing a confining piece, a short distance piece with asymptotic
freedom, plus spin-dependent interactions.

\subsubsection{Hypothesis of saturation by n = 0 states.}
\hspace*{\parindent}
Using this model within the
Bakamjian and Thomas scheme, one finds, for the IW functions of the $n
= 0$ states ($n$ denoting the radial quantum number) \cite{17r}~:
\begin{eqnarray} \label{65} &&\xi (w) = \left ( {2 \over w+1}\right
)^{2\rho^2}\nn \\ &&\tau_{3/2}^{(0)}(w) = \tau_{3/2}^{(0)}(1)\left ( {2
\over w+1}\right )^{2\sigma_{3/2}^2}\nn \\ &&\tau_{1/2}^{(0)}(w) =
\tau_{1/2}^{(0)}(1)\left ( {2 \over w+1}\right )^{2\sigma_{1/2}^2}
\end{eqnarray}

\noi with
\beq \label{66} \rho^2 = 1.02 \eeq

\noi for the slope of the elastic IW function, and, for the transition
IW functions to the lowest $P$-wave states
\begin{eqnarray} \label{67} &\tau_{3/2}^{(0)}(1) = 0.5394 &\qquad
\sigma_{3/2}^2 = 1.50\nn \\ &\tau_{1/2}^{(0)}(1) = 0.2248 &\qquad
\sigma_{1/2}^2 = 0.83\ . \end{eqnarray}

\noi It has been shown that these $n=0$ transition IW functions
dominate the Bjorken \cite{1r}, \cite{2r} and Uraltsev SR \cite{3r}, that read,
respectively
\begin{eqnarray} \label{68} &&\rho^2 = {1 \over 4} + \sum_n
|\tau_{1/2}^{(n)}(1)|^2 + 2 \sum_n  |\tau_{3/2}^{(n)}(1)|^2 \nn \\
&&\nn \\ &&\sum_n |\tau_{3/2}^{(n)}(1)|^2 - \sum_n
|\tau_{1/2}^{(n)}(1)|^2 = {1 \over 4}\ . \end{eqnarray}

\noi Keeping only the $n=0$ states and the numbers quoted above we get
\begin{eqnarray} \label{69} &&{1 \over 4} +
|\tau_{1/2}^{(0)}(1)|^2 + 2 |\tau_{3/2}^{(0)}(1)|^2 = 0.882\nn \\
&&|\tau_{3/2}^{(0)}(1)|^2 -  |\tau_{1/2}^{(0)}(1)|^2 = 0.240  \end{eqnarray}

\noi to be compared, respectively, with $\rho^2 = 1.02$ (\ref{66})
and with 1/4 in the r.h.s. of the second equation (\ref{68}). The $n 
= 0$ states give a
dominant contribution, and saturate
Bjorken SR at the 10 \% level. Uraltsev SR is even more accurate in this
approximation.\par

Let us first test equation (\ref{47e}), saturating it with the $n = 0$
states~:
\beq \label{70} \overline{\Lambda} \cong {1 \over \xi (w) }
\left [ 2(w+1) \Delta E_{3/2}^{(0)} \ \tau_{3/2}^{(0)}(1) \
\tau_{3/2}^{(0)}(w) + \ 2 \Delta E_{1/2}^{(0)}
\ \tau_{1/2}^{(0)}(1) \ \tau_{1/2}^{(0)}(w)\right ] \ . \eeq

Inserting the phenomenological IW functions (\ref{65}) and the values
for $\Delta E_j^{(0)}$ $(j = {1 \over 2}, {3 \over 2})$ obtained in the
same BT scheme \cite{20new}~:
\beq \Delta E_{3/2}^{(0)} \cong \Delta E_{1/2}^{(0)} =
0.406\ {\rm GeV} \eeq

\noi  we get indeed a value of
$\overline{\Lambda}$ that is quite stable in the whole physical region of
$B \to D^*\ell \nu$, $1 \leq w \leq 1.5$. We find
\beq \overline{\Lambda} = 0.513 \pm 0.015\ . \label{71} \eeq

The stability of the result for $\overline{\Lambda}$ is quite
remarkable, and results essentially from the function
$f(w) = (w+1)\tau_{3/2}^{(0)}(w)$ in the r.h.s., that has a slope
$f'(1) = - 1$, very close to the slope of $\xi (w)$, $\xi'(1) = -
1.02$.\par

   On the other hand, one gets, from (\ref{44e}), (\ref{45e}),
(\ref{eqnew}) and the $n=0$ approximation~:
\beq \eta (1) = 0.380 \qquad \qquad \eta '(1) = - 0.006\qquad \qquad
\eta ''(1) =  0.0003\ . \label{72}
\eeq

For the moment, let us notice that the accuracy of the $n = 0$
approximation depends strongly on the considered quantity, at least at
$w=1$. We have seen that in Bjorken and Uraltsev SR the $n = 0$ states
dominate, but the precision is quite different in both cases. As for
the subleading
quantities, defining
\beq \label{79}
R = {\sum\limits_n \Delta
E_{1/2}^{(n)}|\tau_{1/2}^{(n)}(1)|^2 \over \sum\limits_n \Delta
E_{3/2}^{(n)}|\tau_{3/2}^{(n)}(1)|^2}
\eeq

\noi one gets in BT models, keeping only $n = 0$
\beq
R_{BT} = 0.174 \ .
\eeq

\subsubsection{Excited states (n $\not=$ 0) contribution.}
\hspace*{\parindent}
Let us now discuss how this can be modified by $n\not= 0$ states.
Although the results obtained keeping only the $n=0$ states are
encouraging, one must address the question
of the $n \not= 0$ excited states contribution to the
SR. This question can have a clear cut answer within the BT scheme, but
asks for further numerical calculations to compute $\Delta E_j^{(n)}$
and the $w$-dependent form factors $\tau_j^{(n)}(w)$ ($j = {1 \over 2},
{3 \over 2})$, and will be done in a near future.\par

At $w = 1$, we can for the moment state that a sum including higher
$n \not= 0$ states leads, for the quantity (\ref{79}), to the value
\cite{12r}
\beq
\label{equation86}
R_{BT} \cong 0.24
\eeq

\noi giving
\beq
\label{equation87}
\eta (1) \cong 0.34 \ ,
\eeq

\noi that differs from (\ref{72}) by 10~\%. Therefore, it
is of importance to compute the contributions of $n \not= 0$ states for
the different quantities, in particular to $\overline{\Lambda}$ of
which (\ref{71}) is only a lower limit, due to
the positivity of the different contributions. It is also worth
to check whether the inclusion of the $n\not= 0$ states in
(\ref{47e}) yields indeed a constant.\par

In practice, one is anyway confronted to sums truncated to a definite
$n_{max}$. It is of importance to notice that the dependence of the sum
on $n_{max}$ requires the consideration of the radiative corrections.
On the one hand, one approach proposes to identify the renormalization
scale $\mu$ with $\Delta E^{(n_{max})}$ \cite{reference21}. On
the other hand, another point of view \cite{10r}, \cite{11r} distinguishes
between the cut in $n$, given by a scale $\Delta$ such that $\Delta
E^{(n_{max})} \cong \Delta$ and the renormalization point $\mu$,
although eventually both scales can be chosen to be proportional. The
discussion of the contribution of the higher $n$ states is not simple
and cannot be done without including the radiative corrections.

\subsection{Comparison with the QCD Sum Rules approach.}
\hspace*{\parindent}
Although more precise calculations remain to be done in the BT model,
let us qualitatively compare with other approaches. \par

The approach used up to now to obtain information on the subleading
functions has been the QCD Sum Rules (QCDSR) approach \cite{18r} (for
a review, see \cite{19r}). Moreover, radiative
corrections to the subleading $1/m_Q$ corrections have also been computed
within this scheme in these works. For a recent discussion of the
subleading IW functions and their radiative corrections see \cite{20r} and
\cite{11r}. \par

We must first notice that the results obtained from the SR of the
present paper and those of QCDSR are quite different in spirit. In our
approach we have used the BT quark model to compute the r.h.s. of the SR
(\ref{44e}) and (\ref{45e}), while in the QCDSR approach one computes
directly the l.h.s. of these equations.

The subleading IW functions are non-perturbative quantities, and their
calculation within the QCDSR approach is to some extend model-dependent
because it is subject to a number of approximations.\par

Hence the interest of having information on these non-perturbative
quantities within the present method of Bjorken-like SR. We have
considered here only a limited number of quantities, namely the
subleading corrections of the current perturbation type~: $L_4(w)$,
$L_5(w)$ and $L_6(w)$ in the notation of Falk and Neubert \cite{8r}. Moreover,
the radiative corrections to these quantities within the present 
approach have not
been computed.  We must compare the values obtained to the ones of 
the QCDSR without
including radiative corrections. \par

{\it Without including QCD corrections}, the QCDSR method gives the
values \cite{19r}
\bea \label{77} &&\overline{\Lambda} = 0.50 \nn \\ &&\eta (1) = {1
\over 3}\nn \\ &&\eta ' (1) \cong 0
\eea

\noi and sets \cite{11r}
\beq \label{78} \eta '' (1) = 0 \eeq

\noi that are qualitatively consistent with our results (\ref{72}),
(\ref{equation87}).\par

Notice that, as already pointed out in \cite{12r}, the QCDSR algebraic value
$\eta (1) = {1 \over 3}$ would correspond to the value
\beq
R_{QCDSR} = {1 \over 4}
\eeq

\noi that is very close to the value (\ref{equation86}) including
$n\not= 0$ states \cite{reference24}, \cite{25r}.  \par

As we realize in this comparison with the results of QCDSR, we have
only computed in our approach a part of the subleading
non-perturbative corrections, namely the $1/m_Q$ perturbations to the
current. This is a part of a larger program that should include the
subleading quantities related to the
perturbations of the Lagrangian, namely $L_1(w)$, $L_2(w)$ and $L_3(w)$
or, in the more usual phenomenological notation, $\chi_1(w)$,
$\chi_2(w)$ and $\chi_3(w)$ (see, for example \cite{11r} and \cite{20r}).

\section{Conclusion and outlook.} \hspace*{\parindent}
In this paper we have shown that the consideration of the non-forward
amplitude leads to powerful results for the subleading form factors at
order $1/m_Q$, at least in the case of the functions that correspond to
perturbations of the heavy quark current, $L_4(w)$, $L_5(w)$ and
$L_6(w)$, or equivalently $\overline{\Lambda}\xi (w)$ and $\xi_3(w)$,
where $\xi (w)$ is the leading elastic IW function and
$\overline{\Lambda} = m_B - m_b$. \par

The parameters $1/m_b$ and $1/m_c$ are independent,
and we have in this paper studied the sum rules coming from the terms
in $1/m_b$, i.e.
taking $m_c \to \infty$. The method in this case appears somewhat
cumbersome but straightforward. The final results are very simple. We
have considered in the SR intermediate states with orbital angular
momenta of the light quark $\ell = 0, 1$ and 2. The consequences that
we draw from these states made explicit have been easily generalized to all
$\ell$. \par

Within the framework of the OPE and the non-forward amplitude, the
sum rules of the type $B(v_i) \to D^{(n)}(v') \to B(v_f)$ that
depend on the three variables $(w_i,w_f,w_{if}) = (v_i \cdot
v',v_f\cdot v',v_i\cdot v_f)$, allow to write $\overline{\Lambda}\xi
(w)$ and $\xi_3(w)$ in terms of leading quantities, namely the
transition IW functions $\tau_{1/2}^{(n)}(w)$, $\tau_{3/2}^{(n)}(w)$
and the corresponding level spacings $\Delta E_{1/2}^{(n)}$, $\Delta
E_{3/2}^{(n)}$. This has been possible by taking the limit to the
frontier of the domain
$(w_i,w_f,w_{if}) = (w,1,w)$. Then, most of the excited intermediate
states that contribute to the hadronic side of the SR vanish, because
zero recoil is chosen on one side, namely $w_f = 1$, and only the
$P$-wave IW functions $\tau_{1/2}^{(n)}(w)$, $\tau_{3/2}^{(n)}(w)$
survive. As a result, the fundamental quantity of HQET $\overline{\Lambda}$
appears to be a {\it ratio of leading functions} and $\xi_3(w)$ is also
given in terms of leading functions. \par

To proceed further phenomenologically, we have used as an Ansatz for these
functions the results of the Bakamjian-Thomas quark model, that gives
covariant form factors in the heavy quark limit, satisfies IW scaling
and also Bjorken and Uraltsev sum rules. One obtains in this way for a
very wide range of $w$ the expected constancy for $\overline{\Lambda}$,
with a numerical value of the order of 0.5. This value is in agreement
with the QCDSR approach. We find also numerical agreement between our
approach for the ratio of functions $\eta (w) =
\xi_3(w)/\overline{\Lambda}\xi (w)$ and the QCDSR results. It must be
emphasized that the confirmation of the results of the QCDSR approach
by our rigorous method (with the phenomenological input of the BT
model)  is quite encouraging, since both methods are very different in
spirit.\par

The program to study the subleading form
factors presented in this paper should be pursued in several
directions. First, one should also compute the -- in principle
independent -- sum
rules of the $1/m_c$ type ($c$ being the intermediate quark), that
should be consistent with the ones of the $1/m_b$ type computed here.
We safely conjecture that no new SR for the form factors
$\overline{\Lambda}\xi (w)$ and $\xi_3(w)$ will be found. Second, one
should compute within the BT scheme the $n\not= 0$ contributions to
the r.h.s. of the SR (\ref{44e}) and (\ref{45e}). Thirdly, one should
study the other subleading quantities in
the same spirit, namely the form factors that come from perturbations
of the Lagrangian, i.e. $L_1(w)$, $L_2(w)$, $L_3(w)$, that remain
rather uncertain in the QCDSR approach. However, it is not clear that
{\it usable} relations could be obtained in this case, in terms of
computed form factors in the BT class of relativistic quark models. But
this direction should be pursued. Finally, one should follow and
discuss the experimental situation for
the $n = 0$ $P$-wave IW functions $\tau_{1/2}^{(0)}(w)$,
$\tau_{3/2}^{(0)}(w)$. As shown in Appendix~B, the observed
values of $\tau_{1/2}^{(0)}(1)$ do not seem to fit at present
Uraltsev sum rule and the predictions of the BT quark models. However,
the ${1\over 2}^+$ states are very wide, and new semileptonic and
non-leptonic data are necessary to finally settle the question of the
values for $\tau_{1/2}^{(0)}(w)$.

\newpage
\section*{Appendix A.} \hspace*{\parindent} We give here a proof of the
identity between the subleading quantities $\overline{\Sigma}$ defined
by Uraltsev \cite{3r} and $\xi_3(1)$ defined by Falk and Neubert
\cite{8r}
$$\xi_3(1) = \overline{\Sigma} \ . \eqno({\rm A}.1)$$

Uraltsev expands the matrix element between two $B^*$ mesons for
small velocity transfer $\vec{u}$ (formula (14) of the first ref.
\cite{3r}),
$$<B^*(\varepsilon ', \vec{u})|\overline{Q}iD_jQ(0)|B^*(\varepsilon ,
\vec{0})>$$
$$= - {\overline{\Lambda} \over 2} u_j (\varepsilon '^*\cdot
\varepsilon ) + {\overline{\Sigma} \over 2} \left [
(\vec{\varepsilon}\ '^*\cdot \vec{u}) \varepsilon_j - \varepsilon
'^*_j (\vec{\varepsilon} \cdot \vec{u})\right ] + O(\vec{u}^{\
2})\eqno({\rm A}.2)$$

\noi where $Q$ is the heavy quark field. \par

On the other hand, using the formula of Falk and Neubert (3.4)
\cite{8r}, that is valid for all $\Gamma^{\alpha}$
$$<B^*(\varepsilon ', u')|\overline{Q}\Gamma^{\alpha}
iD_{\alpha}Q(0)|B^*(\varepsilon , u)> = - Tr \left [ \xi_{\alpha}
(u,u') \overline{\cal B}^*(\varepsilon ', u') \Gamma^{\alpha} {\cal
B}^*(\varepsilon , u)\right ] \eqno({\rm A}.3)$$

\noi where
$$\xi_{\alpha} (u, u') = \xi_+(w) (u+u')_{\alpha} + \xi_- (w)
(u-u')_{\alpha} - \xi_3(w) \gamma_{\alpha} \eqno({\rm A}.4)$$

\noi one can write, in a covariant way
$$<B^*(\varepsilon ', u)|\overline{Q}iD_jQ(0)|B^*(\varepsilon , v)> =
- Tr \left [ \xi_j (u,v) \overline{\cal B}^*(\varepsilon ', u) {\cal
B}^*(\varepsilon , v)\right ] \ . \eqno({\rm A}.5)$$

Computing in terms of the $\xi_j(u,v)$ the matrix element (A.2) one has
$$<B^*(\varepsilon ', \vec{u})|\overline{Q}iD_jQ(0)|B^*(\varepsilon ,
\vec{0})> \ =  - Tr \left [ \xi_j (u,v) \overline{\cal
B}^*(\varepsilon ', u) {\cal B}^*(\varepsilon , v)\right ]\eqno({\rm
A}.6)$$

\noi with
$$u = \left ( \sqrt{1 + \vec{u}^2}, \vec{u}\right ) \qquad v = (1,
\vec{0}) \ . \eqno({\rm A}.7)$$

Therefore
$$<B^*(\varepsilon ', \vec{u})|\overline{Q}iD_jQ(0)|B^*(\varepsilon ,
\vec{0})> \ = - Tr \left [ \xi_j (u,v) {/\hskip - 2 truemm\varepsilon
}'^* P_+(u) P_+(v) {/\hskip - 2 truemm\varepsilon }\right ]$$
$$= - Tr \left [ \left ( \xi_+(w)  (u+ v)_j + \xi_-(w) (u-v)_j -
\xi_3(w) \gamma_j\right )  {/\hskip - 2 truemm\varepsilon }'^* P_+(u)
P_+(v) {/\hskip - 2 truemm\varepsilon }\right ]\eqno({\rm A}.8)$$

\noi where $w = \sqrt{1 + \vec{u}^{2}}$. \par

Taking into account that {\it under the trace} (A.8) \cite{8r}
$$2\xi_+(1) + \xi_3(1) = 0 \eqno({\rm A}.9)$$

\noi and the relation
$$\xi_-(1) = - {\overline{\Lambda} \over 2} \eqno({\rm A}.10)$$

\noi one obtains, after some algebra, expanding in powers of $\vec{u}$,
$$<B^*(\varepsilon ', \vec{u})|\overline{Q}iD_jQ(0)|B^*(\varepsilon ,
\vec{0})>  $$
$$= - {\overline{\Lambda} \over 2} u_j(\varepsilon '^* \cdot
\varepsilon) + {\xi_3(1) \over 2} \left [ (\vec{\varepsilon}\
'^*\cdot \vec{u})\varepsilon_j - \varepsilon'^*_j(\vec{\varepsilon}
\cdot \vec{u})\right ] + O(\vec{u}^2)\eqno({\rm A}.11)$$

\noi that compared with (A.2) demonstrates the identity (A.1).

\vskip 2 truecm
\section*{Appendix B. Comment on the experimental\break\noindent
situation for $\tau_{\bf 1/2}{\bf (w)}$, $\tau_{\bf 3/2}{\bf (w)}$.}
\hspace*{\parindent}
The aim of the present paper has been a theoretical one. However, a
brief comment on the experimental situation for the states
$D^{**}(j^P,J^P)$ and the corresponding IW functions $\tau_{1/2}(w)$,
$\tau_{3/2}(w)$ is in order.\par

In the framework of the Bakamjian-Thomas quark model, we have given
the prediction of the
shape of the functions $\tau_{1/2}^{(n)}(w)$, $\tau_{3/2}^{(n)}(w)$
for the $n=0$
states. We have seen that these $n = 0$ IW functions almost
saturate the Bjorken and Uraltsev SR, and give quite reasonable results
for the subleading quantities $\overline{\Lambda}\xi (w)$ and
$\xi_3(w)$. \par

However, for the ${1 \over 2}^+$ states, the present experimental
situation seems at odds with these expectations.\par

The more complete experimental analysis, from Belle, reports four
excited states $D^{**}$ above the ground state in non-leptonic decays
$B^- \to D^{**0}\pi^-$ \cite{21r}, two broad and two narrow, that
complete the expected number of $n = 0$ states with parity $P = +$. The
wide states should correspond to the $(j^P,J^P)$ states $\left ( {1
\over 2}^+,0^+\right )$, $\left ( {1 \over 2}^+,1^+\right )$, and the
narrow to $\left ( {3 \over 2}^+,1^+\right )$, $\left ( {3 \over
2}^+,2^+\right )$. In the following, we denote these states $D_J^j$.
Indeed, the strong decays
proceed through $D_0^{1/2}\to D\pi$, $D_1^{1/2}\to D^*\pi$ ($S$-wave),
$D_1^{3/2}\to D^*\pi$, $D_2^{3/2}\to D\pi$, $D^*\pi$
($D$-wave). The product of branching ratios $B(B^- \to
D^{**0}\pi^-)\times B(D^{**0} \to D^{(*)0}\pi^-)$ have been measured.
Assuming factorization of $\pi^-$ emission, assuming also that the
states $1^+$ are unmixed, and
using a simple quark model for elementary pion emission for the decays
$B \to D^{**}\pi$ to estimate needed spin counting coefficients, one
finds qualitatively, as we will see below, a magnitude

$$|\tau_{1/2}^{(0)}(w_0)| \sim |\tau_{3/2}^{(0)}(w_0)| \eqno({\rm B.1})$$

\noi where $w_0 \cong {m_B^2 + m_{D^{**}}^2 \over 2 m_Bm_{D^{**}}}$ is the
value of $w$ for $q^2 = m_{\pi}^2 \cong 0$. This situation is
supported by older
experiments measuring $B \to D^{**}\pi$ or $B \to D^{**}\ell \nu$
\cite{24r}. If the identification of these states is the correct one, the
experimental situation for the states ${1\over 2}^+$ is at odds with
the expectation of the Bakamjian-Thomas quark model. However, an
approximate saturation of Bjorken and Uraltsev SR
by the $n = 0$ states is not excluded within $1\sigma$.\par

However, before concluding about these states, new experimental
confirmation is needed, mostly for the {\it very wide} ${1\over 2}^+$ states.
On the other hand, on the phenomenological side, one should take into
account $1/m_c$ and $1/m_b$ corrections in the decays $B(B
\to D^{**}\pi)$ \cite{9r}, and moreover a sensible theoretical scheme for
the decays $D^{**} \to D^{(*)}\pi$ is needed, including $1/m_c$
corrections. New data on the semileptonic decays $B \to
D^{**}\ell \nu$, that would allow to extract directly the functions
$\tau_{1/2}^{(0)}(w)$, $\tau_{3/2}^{(0)}(w)$ would also be very
welcome. A needed detailed analysis to extract $\tau_{1/2}^{(0)}(w)$,
$\tau_{3/2}^{(0)}(w)$ from present data is beyond the scope of the
present paper. Our aim below is to have only a qualitative estimation. \par

The Belle data on the candidates to the four $P$-wave states $D^{**}$
are the following~:\\

\noi $D^{**}\left ( {3 \over 2}^+, 2^+\right )$
$$M_2^{3/2} = (2460.7 \pm 2.1 \pm 3.1)\ \hbox{MeV}$$
$$\Gamma_2^{3/2} = (46.4 \pm 4.4 \pm 3.1)\ \hbox{MeV}$$
$$B(B^- \to D_2^{3/2\ 0} \pi^-) \times B (D_2^{3/2\ 0} \to D^+\pi^-)
= (3.5 \pm 0.3 \pm 0.5) \times 10^{-4}$$
$$B(B^- \to D_2^{3/2\ 0} \pi^-) \times B (D_2^{3/2\ 0} \to
D^{*+}\pi^-) = (2.0 \pm 0.3 \pm 0.5) \times 10^{-4} \eqno({\rm B.2})$$

\vskip 5 truemm
\noi $D^{**}\left ( {3 \over 2}^+, 1^+\right )$
$$M_1^{3/2} = (2423.9 \pm 1.7 \pm 0.2)\ \hbox{MeV}$$
$$\Gamma_1^{3/2} = (26.7 \pm 3.1 \pm 2.2)\ \hbox{MeV}$$
$$B(B^- \to D_1^{3/2\ 0} \pi^-) \times B (D_1^{3/2\ 0} \to
D^{*+}\pi^-) = (6.2 \pm 0.5 \pm 1.1) \times 10^{-4} \eqno({\rm B.3})$$

\vskip 5 truemm
\noi $D^{**}\left ( {1 \over 2}^+, 0^+\right )$
$$M_0^{1/2} = (2290 \pm 22 \pm 20)\ \hbox{MeV}$$
$$\Gamma_0^{1/2} = (305 \pm 30 \pm 25)\ \hbox{MeV}$$
$$B(B^- \to D_0^{1/2\ 0} \pi^-) \times B (D_0^{1/2\ 0} \to
D^{+}\pi^-) = (5.5 \pm 0.5 \pm 0.8) \times 10^{-4} \eqno({\rm B.4})$$

\vskip 5 truemm
\noi $D^{**}\left ( {1 \over 2}^+, 1^+\right )$
$$M_1^{1/2} = (2400 \pm 30 \pm 20)\ \hbox{MeV}$$
$$\Gamma_1^{1/2} = (380 \pm 100 \pm 100)\ \hbox{MeV}$$
$$B(B^- \to D_1^{1/2\ 0} \pi^-) \times B (D_1^{1/2\ 0} \to
D^{*+}\pi^-) = (4.1 \pm 0.5 \pm 0.8) \times 10^{-4} \ .\eqno({\rm
B.5})$$

Assuming that these states decay essentially into two-body modes, i.e.
$B(D_2^{3/2} \to (D + D^*)\pi)$, $B(D_1^{3/2} \to D^*\pi)$,
$B(D_0^{1/2} \to D\pi)$, $B(D_1^{1/2} \to D^*\pi)$, the following
branching ratios are given by a Clebsch-Gordan coefficient

$$B(D_1^{3/2\ 0 } \to D^{*+}\pi^-) =  B(D_0^{1/2\ 0 } \to D^{+}\pi^-) =
B(D_1^{1/2\ 0 } \to D^{*+}\pi^-) = {2 \over 3}\ . \eqno({\rm B.6})$$

To estimate $B(D_2^{3/2\ 0 } \to D^{+}\pi^-)$ and $B(D_2^{3/2\ 0 } \to
D^{*+}\pi^-)$, we use the spin counting of the non-relativistic quark
model. \par

{\it In the limit of neglecting spin-dependent perturbations} in the
spectrum, i.e. assuming the pairs $(D, D^*)$, $(D_2^{3/2}, 
D_1^{3/2})$ and $(D_1^{1/2}, D_0^{1/2})$ to be degenerate and that 
all the $D_J^j$ are
degenerate, simple angular momentum calculations give, for the {\it
total} widths~:

$$\Gamma (D_2^{3/2}) = \Gamma (D_1^{3/2})  \qquad \qquad \qquad
\Gamma (D_0^{1/2}) =
\Gamma (D_1^{1/2}) \eqno({\rm B.7})$$
$$\Gamma (D_2^{3/2} \to D^*\pi) = {3\over 2} \Gamma (D_2^{3/2} \to D
\pi )\ . \eqno({\rm B.8})$$

\noi This last relation gives the needed spin counting coefficient.
\par

It is easy to obtain this factor by realizing that to have the $D$ wave
(1 denoting the quark emitting a pion and taking
$Oz$ along the pion momentum) one needs the operator
$(\sigma_1^zk_{\pi}) \exp (iz_1 k_{\pi}) \to i k_{\pi}^2 \sigma_1^z
z_1$. We have then, for the non-vanishing amplitudes

$$M(D_2^{3/2} \to D \pi ) =\ <1\ 0, 1\ 0|2\ 0>\ <0\ 0|\sigma_1^z|1\
0>\ <0\ 0|{\cal Y}_1^z|1\ 0>$$
$$M(D_2^{3/2(\pm 1)} \to D^{*(\pm 1)} \pi ) =\ <1\ 0, 1\ \pm 1|2\ \pm
1>\ <1\ \pm 1|\sigma_1^z|1\ \pm 1>\ <0\ 0|{\cal Y}_1^z|1\ 0>
\eqno({\rm B.9})$$

\noi that gives

$$M(D_2^{3/2(\pm 1)} \to D^{*(\pm 1)} \pi ) = \pm {\sqrt{3} \over 2}
\ M(D_2^{3/2} \to D\pi) \eqno({\rm B.10})$$

\noi and hence (B.8). \par

We now take into account the actual masses. Since both $D_2^{3/2} \to
D\pi$ and $D_2^{3/2} \to D^*\pi$ proceed through the $D$-wave, we
will have

$${\Gamma (D_2^{3/2} \to D^*\pi) \over \Gamma (D_2^{3/2} \to D\pi )}
= {3 \over 2} \ {p^{*5} \over p^5} \cong 0.40 \eqno({\rm B.11})$$

\vskip 5 truemm
\noi in an obvious notation. Therefore, we obtain the branching ratios

$$B(D_2^{3/2\ 0} \to D^{+} \pi^- ) \cong 0.48 \qquad B(D_2^{3/2\ 0}
\to D^{*+} \pi^- ) \cong 0.19 \ .\eqno({\rm B.12})$$

\noi From these BR we find, adding the errors in quadrature
$$\begin{array}{ll} B(B^- \to D_2^{3/2\ 0} \pi^- ) = (7.3 \pm 1.2)
\times 10^{-4}  &\hbox{(from $D_2^{3/2\ 0} \to D^+\pi^-$)} \\
B(B^- \to D_2^{3/2\ 0} \pi^- ) = (10.5 \pm 3.1) \times 10^{-4}
&\hbox{(from $D_2^{3/2\ 0} \to D^{*+}\pi^-$)}\ .\end{array}\eqno({\rm
B.13})$$

We realize that the value for $B(B^- \to D_2^{3/2\ 0} \pi^- )$ differs if one
obtains it from $D_2^{3/2\ 0} \to D^+\pi^-$ or from $D_2^{3/2\ 0} \to
D^{*+}\pi^-$, although they agree within $1\sigma$. Using (B.5) for
the other modes, and taking into account the uncertainty from both
results (B.13) one finds
$$B(B^- \to D_2^{3/2\ 0} \pi^- ) = (9.8 \pm 3.8) \times 10^{-4}$$
$$B(B^- \to D_1^{3/2\ 0} \pi^- ) = (9.3 \pm 1.7) \times 10^{-4}$$
$$B(B^- \to D_1^{1/2\ 0} \pi^- ) = (6.1\pm 1.4) \times 10^{-4}$$
$$B(B^- \to D_0^{1/2\ 0} \pi^- ) = (8.2\pm 1.4) \times 10^{-4}
\eqno({\rm B. 14})$$

The decays $B^- \to D^{**0}\pi^-$ proceed through two different
diagrams~: a color-allowed diagram with $\pi^-$ emission, and a
color-suppressed diagram with $D^{**0}$ emission. We will now assume
that the $\pi^-$ emission diagram dominates, although this hypothesis
could be incorrect, as we argue at the end of this appendix. However,
assuming factorization of $\pi^-$ emission and that the states
$1^+$ are unmixed, we find for the decay rates, from \cite{2r}~:

$$\Gamma = {G_F^2 \over 16 \pi} \ |V_{cb}|^2 \ f_{\pi}^2 \ {p \over
m_B^2} \overline{|M(B \to D^{**}\pi )|^2} \ .\eqno({\rm B.15})$$

   $$\overline{|M(B \to D_2^{3/2}\pi )|^2} = 2m_{D^{**}}m_B (m_B +
m_{D^{**}})^2 (w_0^2 - 1)^2 |\tau_{3/2}(w_0)|^2$$
$$\overline{|M(B \to D_1^{3/2}\pi )|^2} = 2m_{D^{**}}m_B (m_B -
m_{D^{**}})^2 (w_0 + 1)^2 (w_0^2 - 1) |\tau_{3/2}(w_0)|^2$$
$$\overline{|M(B \to D_1^{1/2}\pi )|^2} = 4m_{D^{**}}m_B (m_B -
m_{D^{**}})^2 (w_0^2 - 1) |\tau_{1/2}(w_0)|^2$$
$$\overline{|M(B \to D_0^{1/2}\pi )|^2} = 4m_{D^{**}}m_B (m_B +
m_{D^{**}})^2 (w_0 - 1)^2 |\tau_{1/2}(w_0)|^2 \eqno({\rm B.16})$$

\noi with
$$w_0 \cong {m_B^2 + m_{D^{**}}^2 \over 2m_Bm_{D^{**}}} \qquad \qquad
p \cong {m_B^2 - m_{D^{**}}^2 \over 2m_B}\eqno({\rm B.17})$$

\noi the subindex 0 denoting the value of $w$ for $q^2 = m_{\pi}^2
\cong 0$, and $m_{D^{**}}$ the mass of the corresponding $D(j^P,J^P)$
state. \par

It is interesting to notice that the rates (B.15), (B.16) are given
by the expressions
$$\Gamma (B \to D_2^{3/2} \pi ) = \Gamma (B \to D_1^{3/2}\pi ) = \\
{G_F^2 \over 16 \pi} |V_{cb}|^2 m_B^3 \ f_{\pi}^2 {(1 - r)^5 (1+r)^7
\over 16r^3}\left |\tau_{3/2}\left ({1+r^2\over 2r} \right )\right 
|^2 \eqno({\rm B.18})$$
$$\Gamma (B \to D_1^{1/2} \pi ) = \Gamma (B \to D_0^{1/2}\pi ) = \\
{G_F^2 \over 16 \pi} |V_{cb}|^2 m_B^3 \ f_{\pi}^2 {(1 - r)^5 (1+r)^3
\over 2r}\left |\tau_{1/2}\left ({1+r^2\over 2r} \right )\right |^2 
\eqno({\rm B.19})$$

\noi where $r = {m_{D(3/2)} \over m_B}$ and $r = {m_{D(1/2)} \over m_B}$
respectively in the first and the second relations. The equalities
$\Gamma (B \to D_2^{3/2} \pi ) = \Gamma (B \to D_1^{3/2} \pi )$ and
$\Gamma (B \to D_1^{1/2}\pi ) = \Gamma (B \to D_0^{1/2}\pi)$ follow
from heavy quark symmetry, since there is a single helicity amplitude 
in all decays.

Using the central values for the masses, but taking into account
the errors in (B.14), we find respectively for the different
modes, roughly~:

$$\begin{array}{ll} B \to D_2^{3/2}\pi &\qquad \qquad
|\tau_{3/2}(1.30)| = 0.29 \pm 0.06 \\
B \to D_1^{3/2}\pi &\qquad \qquad |\tau_{3/2}(1.32)| = 0.27 \pm 0.04 \\
   B \to D_1^{1/2}\pi &\qquad \qquad  |\tau_{1/2}(1.33)| = 0.35 \pm 0.04\\
B \to D_0^{1/2}\pi &\qquad \qquad |\tau_{1/2}(1.37)| = 0.39 \pm 0.06
\end{array} \eqno({\rm B.20})$$

Within $1\sigma$ there is consistency between the different
determinations of $|\tau_{3/2}(w_0)|$ and $|\tau_{1/2}(w_0)|$, but
errors increase considering both determinations. We conclude safely
that we will have the numbers
$$|\tau_{3/2}(1.31)| = 0.28 \pm 0.06$$
$$|\tau_{1/2}(1.35)| = 0.38 \pm 0.07 \eqno({\rm B.21})$$

Extrapolating now with the Bakamjian-Thomas form factors (\ref{65}),
(\ref{67}) for $\tau_{3/2}(w)$ and $\tau_{1/2}(w)$, we get
$$\begin{array}{l} |\tau_{3/2}(1)| = 0.44 \pm 0.10 \\ |\tau_{1/2}(1)|
= 0.49 \pm 0.10\end{array} \eqno({\rm B.22})$$

\noi to be compared with the values in the BT model
$|\tau_{3/2}(1)|^{BT} = 0.54$, $|\tau_{1/2}(1)|^{BT} = 0.22$. We find
agreement for $|\tau_{3/2}(1)|$  within errors, but $|\tau_{1/2}(1)|$
is much too large compared with the BT model.\par

Saturating Bjorken and Uraltsev SR with the $n = 0$ states, we get a
contribution to the Bjorken and Uraltsev SR that lies, within
$1\sigma$, in the following range, keeping only the $n=0$ states~:
$$\rho^2 \cong {1 \over 4} + |\tau_{1/2}(1)|^2 + 2 |\tau_{3/2}(1)|^2
= 0.90 \pm 0.28$$
$$|\tau_{3/2}(1)|^2 - |\tau_{1/2}(1)|^2 = - 0.04 \pm 0.18 \ 
.\eqno({\rm B.23})$$

\noi Therefore, within $1\sigma$, low values for $\rho^2$ are not
excluded but the second value is too small compared to the r.h.s. ${1 
\over 4}$ of Uraltsev SR (\ref{68}).
\par

We must keep in mind however that the estimation (B.22), that relies on
the simple hypothesis of the dominance of $\pi^-$ emission could be
incorrect owing to two facts. First, in many decay modes the 
color-suppressed diagrams are
empirically not so suppressed. Second,
the diagram of $D^{**0}$ emission is not computable on the ground of
first principles in the BBNS QCD factorization scheme \cite{23r}, since
the emitted meson is composed of heavy-light quarks. Therefore, one
must keep in mind that new data on semileptonic decays $B \to
D^{**}\ell \nu$ (where the statistics is much smaller than in
non-leptonic decays), that directly measure the functions
$\tau_{1/2}(w)$
and $\tau_{3/2}(w)$, are necessary to settle the question of the
magnitude of $\tau_{1/2}(1)$, $\tau_{3/2}(1)$ and their comparison with
Bjorken and Uraltsev SR and with the predictions of Bakamjian-Thomas models.

\section*{Acknowledgements}

We are indebted to the EC contract HPRN-CT-2002-00311 (EURIDICE) for its
support. One of us (L. O.) acknowledges also hospitality and partial
support from INFN, Milano.

 \end{document}